\documentclass[prd,byrevtex
,preprint
,showkeys
,showpacs
,amsfonts,amsmath,amssymb
]{revtex4}
\usepackage[pdftex]{graphicx}
\usepackage{float}
\usepackage{color}
\usepackage{physics}
\newcommand{\anno}[1]{\textcolor{black}{#1}}
\begin{document}
\title{Quantum Circuit Model of Black Hole Evaporation}
\author{Tomoro Tokusumi}
\email{tokusumi.tomoro@b.mbox.nagoya-u.ac.jp}
\author{Akira Matsumura}
\email{matsumura.akira@h.mbox.nagoya-u.ac.jp}
\author{Yasusada Nambu}
\email{nambu@gravity.phys.nagoya-u.ac.jp}
\affiliation{Department of Physics, Graduate School of Science, Nagoya
University, Chikusa, Nagoya 464-8602, Japan}

\date{October 1, 2018} 
%
\begin{abstract}
  We consider a quantum circuit model describing the evaporation
  process of black holes. We specifically examine the behavior of the
  multipartite entanglement represented by this model, and find that
  the entanglement structure depends on the black hole mass $M$ and
  the frequency of the Hawking radiation $\omega$. For sufficiently
  small values of $M\omega$, the black hole and the radiation system
  becomes a separable state after the Page time and a firewall-like
  structure appears.  On the contrary, for larger values of $M\omega$,
  the entanglement between the black hole and the radiation is not
  lost. These behaviors imply that owing to the monogamy property of
  the multipartite entanglement, low frequency modes of the Hawking
  radiation destroys the quantum correlation between the black hole
  and the emitted Hawking radiation.
\end{abstract}
\maketitle

\tableofcontents

\section{Introduction}
By investigating the quantum field in black hole spacetimes, Hawking
predicted that the black hole spontaneously creates entangled particle
pairs in the vicinity of the black hole
horizon~\cite{Hawking_particle_creation}. One particle of the created
pair is thermally radiated away to the spatially far region, and the
other particle with negative energy falls into the black hole. As the
result, the horizon shrinks and the black hole evaporates. In 1976,
Hawking pointed out that the information paradox exists in the
evaporation process of the black
hole~\cite{Hawking_info_loss}. Assuming that the initial state of the
black hole is pure and keeping in mind that the thermal radiation is
expressed as the mixed state such as the Gibbs state, it turns out
that the black hole becomes a mixed state as the evaporation proceeds.
However, in theories such as the quantum mechanics, the quantum field
theory and the string theory, time evolution is unitary. Thus a pure
state evolves with  unitary operators and the state remains as a pure
state after the evaporation.  Consequently, evaporation process of the
black hole seems to be inconsistent with the quantum theory.

This inconsistency is also shown by considering the entanglement.  The
amount of the entanglement between systems A and B can be evaluated by
the von Neumann entropy of the reduced density operator
$S_\text{A} = -\Tr \rho_\text{A} \log \rho_\text{A}$.  The von Neumann
entropy $ S_\text{A} $ is zero if and only if the system A is in pure
state. In other words, by measuring the amount of entanglement, we can
investigate the presence or absence of information paradox because the
existence of entanglement is connected to mixedness of the considering
system. According to Hawking's calculation, the amount of the
entanglement between a particle falling into the black hole and a
particle radiated away is zero at the initial stage of evaporation and
monotonically increases as the evaporation proceeds. In other words,
particles radiated away from the black hole are in pure states at the
beginning, then become mixed after evaporation completes. This is why
Hawking concluded the information paradox exists.

For this paradox to be solved, the von Neumann entropy should begin to
decrease in the middle stage of the evaporation and finally it must
become zero. Assuming that the black hole and the Hawking radiation
are in a random state and the dynamics of the black hole is chaotic,
Page showed that the von Neumann entropy of the Hawking radiation
increases from zero, and it has a maximum value at the middle stage of
the evaporation and then decreases monotonously. This behavior of the
von Neuman entropy is called the Page curve~
\cite{Page_curve,page_theorem} and the time attaining the maximum of the
entropy is called the Page time.  It is believed that the Page curve
represents the general feature of the evaporation process of the black
hole and the information of the black hole is carried away by the
Hawking radiation.

In studies of AdS/CFT correspondences \cite{AdSCFT}, it is partially
shown that the information paradox of the black hole evaporation does
not exist if  the assumption of the black hole
complementarity~\cite{BH_complementarity} is imposed:
\begin{description}
  \item[Postulate 1:] The process of formation and evaporation of a
  black hole, as viewed by a distant observer, can be described
  entirely within the context of standard quantum theory. In
  particular, there exists an unitary S-matrix which describes the
  evolution from infalling matter to outgoing Hawking-like radiation.
  \item[Postulate 2:] Outside the stretched horizon of a massive black
  hole, physics can be described to good approximation by a set of
  semi-classical field equations.
  \item[Postulate 3:] To a distant observer, a black hole appears to
  be a quantum system with discrete energy levels. The dimension of
  the  states of a black hole with mass  $M$ is the
  exponential of the Bekenstein-Hawking entropy
\begin{equation}
  \label{eq:BHent}
   S_\text{BH}(M)=\frac{\mathcal{A}}{4G},\quad G=m_\text{pl}^{-2},
\end{equation}
where $\mathcal{A}$ is the surface area of the black hole.
\end{description}
Hence, it is widely believed that there is no information paradox with
these assumptions, and a consistent evaporation process occurs in the 
quantum theory. However, considering the black hole after the Page
time, A. Almheiri, D. Marolf, J. Polchinski and J. Sully
(AMPS)~\cite{AMPS} argued that the postulates of the
black hole complementarity are still inconsistent and the
  following postulate should be added:
\begin{description}
  \item[Postulate 4:] A freely falling observer experiences nothing
  out of the ordinary when crossing the horizon.
\end{description}
To make these four postulates consistent, as the most conservative
solution, an observer falling through the horizon burns out before it
passes through. Therefore, AMPS proposed that the horizon is covered
by a firewall.  The ``monogamy'' relation which is the property of
multipartite entanglement plays an important role to show this
inconsistency. The monogamy relation states that one system can not
entangle with any other system when it is already entangled strongly
with another system.  Let us divide a system of the evaporation
process into three subsystems as follows: \textsf{ER} (early
radiation) is a previously emitted Hawking particle, \textsf{JR} is a Hawking
particle just emitted to exterior, \textsf{BH} is a partner of the
Hawking radiation just created and falls into the black hole.  After the
Page time (when the Page curve is decreasing), \textsf{JR} is
entangled strongly with \textsf{ER} due to Page's theorem, and
\textsf{JR} is also entangled with \textsf{BH} due to purity of the
Hawking radiation.  Owing to the postulate 4, there is no drama at the
black hole horizon, so there must be very strong entanglement between
\textsf{JR} and \textsf{BH}. If the monogamy relation holds, these
appears inconsistency. In order to resolve this inconsistency, the
firewall must be introduced to cut the entanglement between
\textsf{JR} and \textsf{BH}.

To justify the existence of the firewall, it is important to
investigate the feature of the multipartite entanglement between the
radiated Hawking particles and the particles falling to the black
hole. In particular, it is necessary to show whether entanglement is
really broken or not. However, it is not easy to evaluate multipartite
entanglement in the context of the quantum field theory and there are
several works investigating more general quantum information structure
by modeling the evaporation process of black holes with qubit
systems~\cite{Mathur:correlation,Braunstein_qubit1,Braunstein_qubit2,Hwang,Albrecht_multi}. A
burning paper model is well known as a physical model to explain the
Page curve~\cite{Mathur:correlation}. When a paper is burning, photons
are emitted from the surface of the paper. The photon can be entangled
with the spin of atoms composing the paper.  The entanglement entropy
between photons and the paper monotonically increases at the beginning.
Then, atoms entangled with photon also become ash and are
emitted to the outside and the entanglement entropy may start
to decrease at a certain time. When all the paper burns out, ash is
released and the entanglement entropy becomes zero. Due to the very
complex interaction of the atom with the emitted photons during the
burning process, the paper's information is fully encoded into
radiated photons and ashes. Although the burning paper model is
adequate to explain the Page curve, it has no horizon
structure. This is because atoms constituting the paper are
discharged to the outside in the form of ashes. Since the horizon is
an important structure of spacetimes to characterize the black hole
\anno{and its size characterizes strength of the entanglement between
  created Hawking pairs}, 
this difference is unacceptable to investigate the evaporation of the
black hole using a toy models with qubits.

Qubit models have been applied to various evaporation
scenarios. J.~Hwang \textit{et al.}~\cite{Hwang} investigated a qubit
model from the viewpoint of the remnant scenarios which are based on a
different postulate from AMPS's one, and revealed restrictions of
various scenarios of remnant picture.  In the research by S.~Luo
\textit{et al.}~\cite{Albrecht_multi}, properties of multipartite
entanglement in the evaporation process were investigated. During the
entire history of evaporation, they found entanglement between
\textsf{BH} and \textsf{JR} differs from AMPS's expectation and they
pointed out the firewall may not be necessary to resolve the
information loss problem . However, the time evolution of evaporation
in their model is different from that of the Hawking radiation.

In this paper, we propose a quantum circuit model with qubits
representing the Hawking radiation with the horizon structure, and
investigate behavior of the multipartite entanglement. We aim to
discuss emergence of the firewall in our model.  The structure of this
paper is as follows. In Sec.~II we introduce our quantum circuit model
representing the black hole evaporation. In Sec.~III, we present results of
numerical simulation of our model. In Sec.~IV, we analytically
  obtain the state after the Page time and investigate the mutual
  information and the negativity between the black hole and the
  Hawking radiation. Sec.~V is devoted to summary and discussion.

\section{Quantum Circuit  Model}
We introduce our quantum circuit model of the black hole  evaporation.
\subsection{Setup}
\begin{figure}[H]
    \centering
    \includegraphics[width=0.8\hsize,clip]{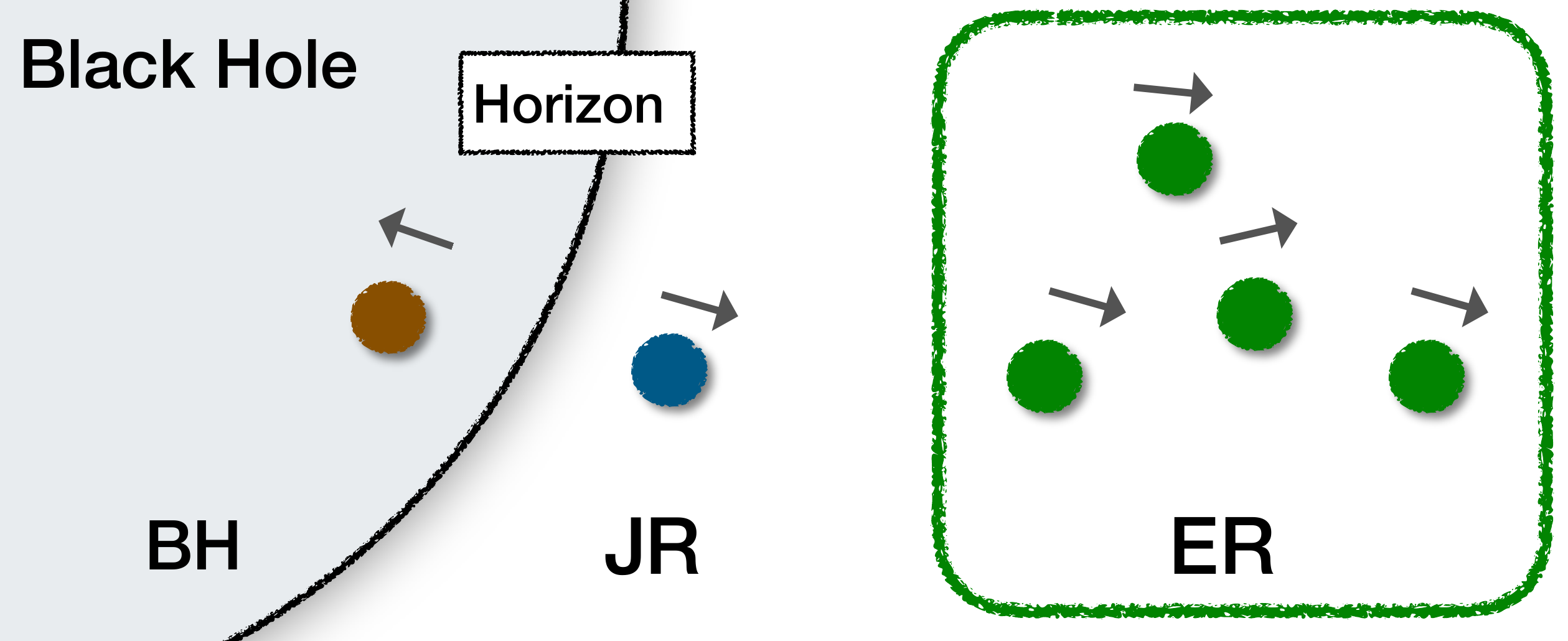}
    \caption{Defition of \textsf{BH}, \textsf{JR} and
        \textsf{ER} in our model.}
    \label{suffix}
 \end{figure}

We describe the Hawking radiation using qubits in our model. A qubit has
two distinct internal states $ \ket{0}$ and $ \ket{1} $. In order to
formulate the black hole evaporation, we separate the Hilbert
space of the total system to that of the black hole, the Hawking
radiation and the Hawking radiation that had been radiated
previously. Hence, prepare the Hilbert space of black hole (\textsf{BH}), just
radiation (\textsf{JR}), early radiation (\textsf{ER}) (Fig.~\ref{suffix})
\begin{align}
  \mathcal{H}_\text{tot}
= \mathcal{H}_\textsf{BH}\otimes \mathcal{H}_\textsf{JR}\otimes\mathcal{H}_\textsf{ER}.
\end{align}
Here, we regard the state $ \ket{0} $ of \textsf{JR} and \textsf{ER}
as the ``vacuum state'' with no particles.
\noindent
We treat particles in \textsf{BH}, that are paired with the
  emitted Hawking particles and fall into the black hole, constitute
  the black hole's degrees of freedom. This is because particles
  falling into the black hole can be regarded as a part of the black
  hole owing to the black hole complementary postulate 4.

\subsection{Evaporation}\label{evaporation}
\begin{figure}[H]
   \centering
   \includegraphics[width=0.2\hsize,clip]{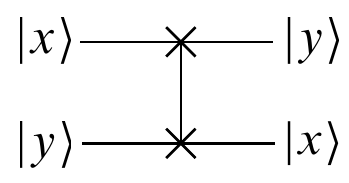}
   \caption{The SWAP gate.}
   \label{fig:SWAP}
 \end{figure}

Particles come out from the black hole in the evaporation
process. For example, a particle in \textsf{JR} will belong to
  \textsf{ER} at the next time step. This process can be realized
using the SWAP operation, which exchanges inputted two qubits and is
represented by the unitary quantum gate shown in Fig.~\ref{fig:SWAP}.
\noindent
In the evaporation process, the black hole shrinks by emitting
particles and finally disappears. Furthermore, the information of the
black hole is transferred to  radiated particles.  Qubits in \textsf{BH}
gradually change to $\ket{0}$ state during the evaporation process. When the
evaporation finishes and the black hole disappears completely, all the
qubits in \textsf{BH} become $\ket{0}$ state. On the other hand, at
the initial stage of evaporation, most of the radiated particles are
 in $\ket{0}$ state (no particles). Thus, radiated
particles state will becomes very complicated because
information of the black hole is carried away by the radiated
particles.  We expect that evaporation can be modeled by swapping
the information (qubit) of the black hole with a part of radiation's
degrees of freedom. In our model, the SWAP are performed twice
between \textsf{BH} and \textsf{JR}, and between \textsf{JR} and \textsf{ER}.

\begin{figure}[H]
   \centering
    \includegraphics[width=0.4\hsize,clip]{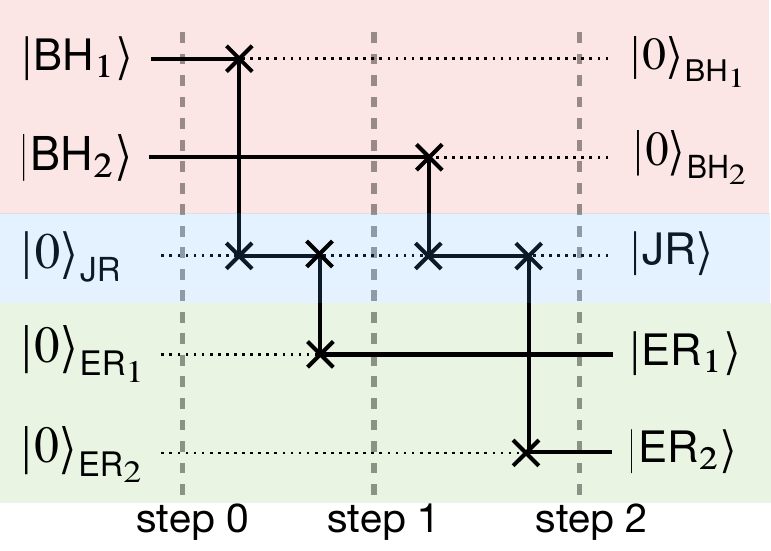}
    \caption{The SWAP circuit for 5 qubit system.} 
   \label{fig:SWAP-circuit}
 \end{figure}

As a demonstration, we look at the case of the 5 qubit
total system and show how the SWAP procedure transfer the information
of \textsf{BH} to radiations.  The total system is
\begin{align}
  \ket{\psi}=\ket{\textsf{BH}}\otimes \ket{\textsf{JR}}\otimes \ket{\textsf{ER}},
\end{align}
with $|\textsf{BH}|=2^2$ (2 qubit), $|\textsf{JR}|=2^1$ (1 qubit),
$|\textsf{ER}|=2^2$ (2 qubit). The initial state of
each subsystem is assumed to be

\begin{equation}
  \anno{\ket{\textsf{BH}}=\frac{\ket{0}_{\textsf{BH}_1}\ket{1}_{\textsf{BH}_2}+\ket{1}_{\textsf{BH}_1}\ket{0}_{\textsf{BH}_2}}{\sqrt{2}}},\quad
   \ket{\textsf{JR}}=\ket{0}_\textsf{JR},\quad
    \ket{\textsf{ER}}=\ket{0}_{\textsf{ER}_1}\ket{0}_{\textsf{ER}_2},
\end{equation}
and the  initial total state $\ket{\psi_0}$ is
\begin{align}
  \ket{\psi_0}=\anno{\frac{1}{\sqrt{2}}}\Bigl[\ket{10}+\ket{01}\Bigr]_{\textsf{BH}_{12}}\ket{0}_\textsf{JR}\ket{00}_{\textsf{ER}_{12}}.
\end{align} 
Although the Page curve is the average of entanglement entropy, we
  focus here on  a specific state and compute the
  non-averaged entanglement entropy. The quantum circuit we introduce
is shown in Fig.~\ref{fig:SWAP-circuit}.
\noindent
Let us follow the time steps realized by this circuit.

\begin{figure}[H]
   \centering
   \includegraphics[width=0.45\hsize,clip]{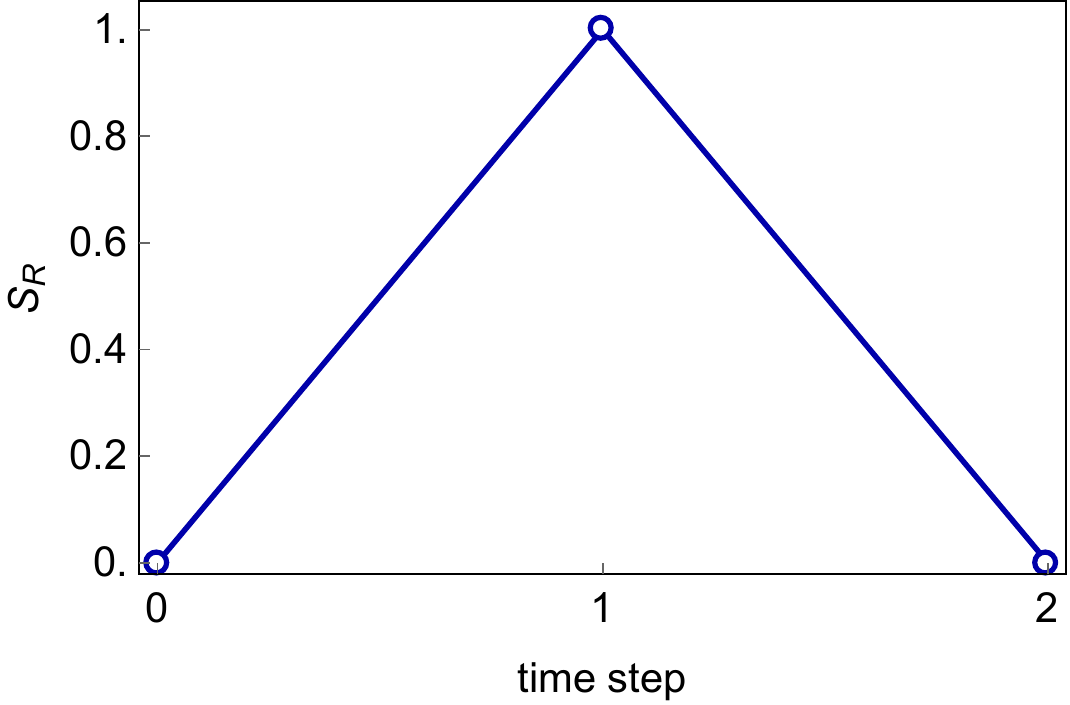}
   \caption{The Page curve obtained by the SWAP circuit in
     Fig.~\ref{fig:SWAP-circuit}.}
   \label{fig:swap0}
 \end{figure}

\noindent
\textsf{At time step 1:}
\begin{enumerate}
\item Apply the SWAP between the first qubit of \textsf{BH} and
  \textsf{JR} ,
        \begin{align}
          \ket{\psi_0}\longmapsto\ket{\psi_{0'}}=\anno{\frac{1}{\sqrt{2}}}\ket{0}_{\textsf{BH}_1}\Bigl[\ket{01}+\ket{10}\Bigr]_{\textsf{BH}_2,\textsf{JR}}\ket{00}_{\textsf{ER}_{12}}.
        \end{align}
\item Apply the SWAP between \textsf{JR} and the first qubit of \textsf{ER},
\begin{align}
  \ket{\psi_{0'}}\longmapsto\ket{\psi_{1}}=\anno{\frac{1}{\sqrt{2}}}\ket{0}_{\textsf{BH}_1}\ket{0}_\textsf{JR}\Bigl[\ket{01}+\ket{10}\Bigr]_{\textsf{BH}_2,\textsf{ER}_1}\ket{0}_{\textsf{ER}_2}.
        \end{align}
\end{enumerate}
\textsf{At time step 2:}
\begin{enumerate}
\item Apply the SWAP between the second qubit of \textsf{BH} and \textsf{JR},
\begin{align}
  \ket{\psi_1}\longmapsto\ket{\psi_{1'}}=\anno{\frac{1}{\sqrt{2}}}\ket{00}_{\textsf{BH}_{12}}\Bigl[\ket{01}+\ket{10}\Bigr]_{\textsf{JR},\textsf{ER}_1}\ket{0}_{\textsf{ER}_2}.
        \end{align}
\item Apply the SWAP between \textsf{JR} and the second qubit of \textsf{ER},
\begin{align}
  \ket{\psi_{1'}}\longmapsto\ket{\psi_{2}}=\anno{\frac{1}{\sqrt{2}}}\ket{00}_{\textsf{BH}_{12}}\ket{0}_\textsf{JR}\Bigl[\ket{10}
+\ket{01}\Bigr]_{\textsf{ER}_{12}}.
\end{align}
\end{enumerate}
\noindent
To check this process reflects the property of evaporating models, we
compute the von Neumann entropy of radiations
\textsf{R}$:=$\textsf{JR}\,$\cup$\,\textsf{ER},
$S_\textsf{R} = - \Tr \rho_\textsf{R} \log_2
\rho_\textsf{R}$ at each step of the SWAP (Fig.~\ref{fig:swap0}).
\noindent
  The typical shape of the Page curve is obtained and the
  information of \textsf{BH} is transferred to the radiation by the SWAP
  operation.  The effective size of \textsf{BH}'s Hilbert space at
  step $n$ is defined by the amount of information
 \begin{equation}
 |\textsf{BH}_n^{\text{eff}}| := |\textsf{BH}|*2^{-(n-1)}\quad (n\geq
 1),\qquad |\textsf{BH}_0^{\text{eff}}| := |\textsf{BH}|
 \end{equation}
 where $|\textsf{BH}|$ is the dimension of \textsf{BH}'s Hilbert space
 determined by the number of qubits.  Particles are radiated from
 \textsf{BH} by the first SWAP at each step. Thus, this quantum
 circuit does not have the black hole horizon and the
 entanglement structure of this circuit is determined only by the
 \textsf{BH} initial state. Qubits in \textsf{BH} are entangled with
 each other at the initial step. Then, a part of entangled qubit is
 moved to become \textsf{JR} and the entanglement between \textsf{BH}
 and \textsf{R} increase. However, the entanglement of the Hawking
 radiation is originated from the entangled particle pairs created
 in the vicinity of the black hole horizon and this SWAP
 circuit is the model of evaporation process without the horizon.

\subsection{Hawking radiation}
\begin{figure}[H]
     \centering
     \includegraphics[width=0.7\hsize,clip]{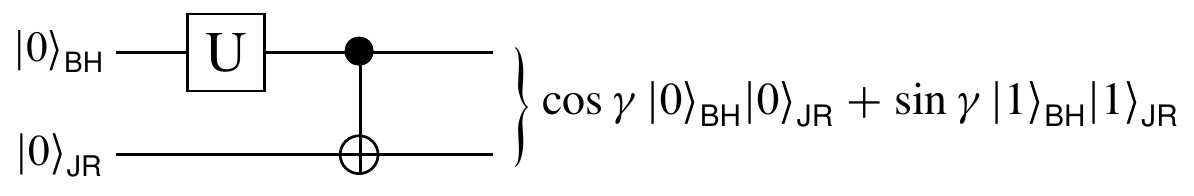}
     \caption{The CNOT-U gate.}
     \label{fig:squeezing-gate} 
  \end{figure}

The quantum state of the Hawking radiation is expressed as
$\prod_{i}\sum_{n_i=0}^\infty e^{-n_i\beta \omega_i/2}
\ket{n_i}_\textsf{BH}\otimes\ket{n_i}_\textsf{JR}$
where $\beta=8\pi M$ is the inverse temperature of the black hole and
$ \omega $ is the frequency of the
radiation~\cite{Hawking_particle_creation}.  The state
$\ket{n_i}_\textsf{BH}\otimes\ket{n_i}_\textsf{JR}$ represents pairs
of the Hawking particles.  For $M\omega\gg 0.1 $, the state is almost
same as the vacuum state without particle excitation and for
$ M\omega < 0.1$, $\ket{1}_\textsf{BH}\otimes\ket{1}_\textsf{JR}$
particle state mainly contributes to the total state of the Hawking
radiation.  Therefore, the essence of the entanglement structure can
be revealed even if two or more particle number states are ignored.
Thus, we assume the following entangled state represents the Hawking
radiation
 \begin{align}
    \frac{1}{\sqrt{1+\exp(-8\pi
   M\omega)}}\biggl[\ket{0}_{\textsf{BH}}\ket{0}_\textsf{JR} 
   +\exp(-4\pi
   M\omega)\ket{1}_{\textsf{BH}}\ket{1}_\textsf{JR}\biggr]. 
\label{eq:haw}
 \end{align} 
 From now on, we focus on a specific $\omega$ mode of the Hawking radiation.
 We consider a quantum gate mimicking this state
 (Fig.~\ref{fig:squeezing-gate}). The gate consists of two well-known
 quantum gates and outputs the entangled state.
\noindent
The gate of the right part in the circuit is an
unitary gate acting on two qubit called the CNOT gate and works as
$\ket{x} \ket{y} \to \ket{x} \ket{x\oplus y}$.
The gate U in the left part is an unitary gate acting on a single
qubit. In terms of the matrix representation, it is
expressed as
\begin{align}
  U := \begin{bmatrix}
       \cos{\gamma} & \sin{\gamma}\\
        \sin{\gamma} & - \cos{\gamma} \label{rotation}
 \end{bmatrix},\quad \tan\gamma=\exp(-4\pi M\omega),
\end{align}
where $\gamma$ is the squeezing parameter.  For $\gamma = \pi/4$, U
corresponds to the Hadamard gate H, and the CNOT-U gate will output
the maximally entangled EPR state. The gate in
Fig.~\ref{fig:squeezing-gate} can output an entangled state and the
amount of entanglement depends on the parameter $ \gamma $.

The strength of the entanglement between the created Hawking particle pairs is
determined by the mass of the black hole \anno{at the considering time
step} and we must determine
the relation between the parameter $ \gamma $ and the black hole
mass. For this purpose, we consider the correspondence between the
dimension of the \textsf{BH} Hilbert space (\textsf{BH} qubit number)
and the black hole mass $ M $.  From the formula of the
Bekenstein-Hawking entropy \eqref{eq:BHent}
\begin{equation}
    S_\text{BH}(M)=4\pi \left(\frac{M}{m_\text{pl}}\right)^2 = \log_2 |\textsf{BH}^\text{eff}|,
\end{equation}
we have
 \begin{align}
   M_n = \frac{m_\text{pl}}{2}\sqrt{\frac{\log_2{|\textsf{BH}_n^\text{eff}|}}{\pi}}.
 \end{align}
 Note that $ | \textsf{BH}_n^\text{eff} | $ is just the effective size
 of \textsf{BH} Hilbert space and not the actual number of the
 \textsf{BH} states.  Using these relations, in the natural unit, we
 obtain
\begin{align}
  \gamma_n = \mathrm{Arctan}\left[\exp\left(-2\omega \sqrt{\pi \log_2
  |\textsf{BH}_n^\text{eff}|}\right)\right].   \label{gamma}
\end{align}
With this squeezing parameter $ \gamma_n $, the CNOT-U gate in
Fig.~\ref{fig:squeezing-gate} mimics the the Hawking radiation (\ref{eq:haw}).

We combine the SWAP gate and the CNOT-U gate to construct a model of
the evaporation process of the Hawking radiation. We have already
confirmed that the initial state of \textsf{BH} is transferred to
\textsf{JR} and \textsf{ER} by the SWAP circuit and the Page curve is
obtained. In the following analysis, we focus on the property of the
entanglement caused by the Hawking radiation to clarify how the
dynamics of the evaporation is connected with the quantum information
process and the firewall argument. Hence, the initial state is assumed
to be non-entangled state
\begin{equation}
\ket{\text{init}} = \ket{000 \cdots 000}_\textsf{BH} \ket{0}_\textsf{JR}
\ket{000 \cdots  000}_\textsf{ER}.
\end{equation}
The quantum circuit of our model is presented in
Fig.~\ref{fig,haw-circuit}.
\begin{figure}[H]
   \centering
   \includegraphics[width=\hsize,clip]{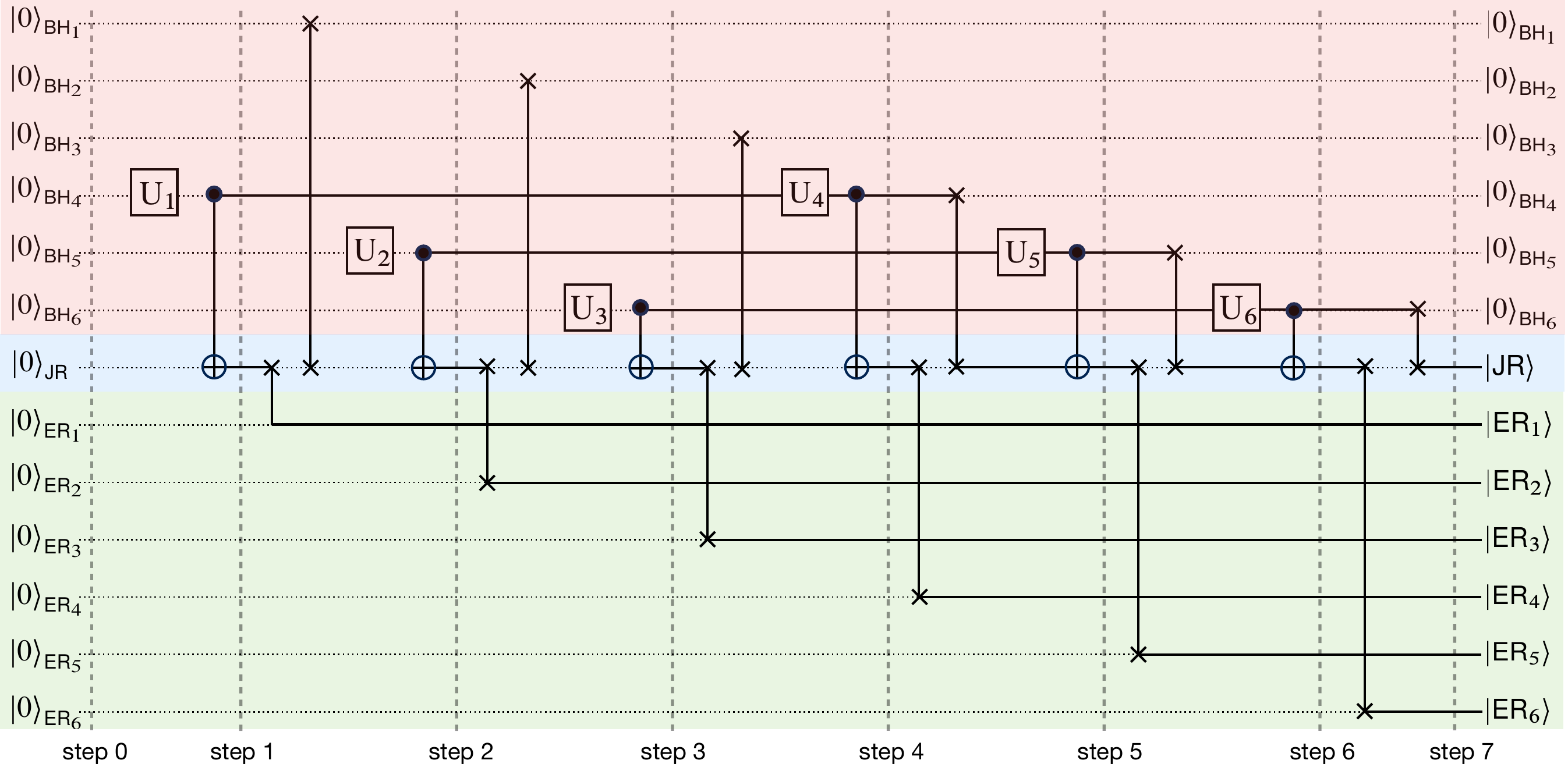}
   \caption{A quantum circuit of the black hole evaporation.  The CNOT-U
     gate acts on \textsf{BH} and \textsf{JR} to generate a Hawking
     particle pair. The SWAP gate acts on two places between
     \textsf{JR} and \textsf{ER}, \textsf{BH} and \textsf{JR} to move
     a qubit of \textsf{JR} to \textsf{ER} and move a qubit in
     \textsf{BH} to \textsf{JR}. We note that the SWAP step is just
     after creation of the Hawking pair, so particles do not directly
     escape from \textsf{BH} to \textsf{JR}.}
   \label{fig,haw-circuit}
\end{figure}

In this model, the initial mass of the black hole corresponds to qubit
number of \textsf{BH}. The qubit number can not make so large due to increase
of the simulation time. 
We denote the black hole mass at a certain step $n$ as
$M_{n} \omega$.  Because $M_n\propto \sqrt{\log_2 |\textsf{BH}_n|}$,
the black hole mass at arbitrary step $n$ is
\begin{align}
     M_n =
  M_{\text{init}}\sqrt{\frac{\log_2|\textsf{BH}_n|}{\log_2|\textsf{BH}_{\text{init}}|}}
 = M_{\text{init}} \sqrt{1-\frac{n-1}{N_\textsf{BH}}},
   \end{align}
   where $M_\text{init}$ is the initial mass of the black hole and
   $N_\textsf{BH}$ is the qubit number of \textsf{BH} used for
   analysis. The only parameters contained in this model are the value
   of $ M_{n} \omega $ and $n$. In our analysis, we choose $n$
   as the step at the Page time.
\section{Numerical Result}
In this section, we will analyze evolution of the state in our circuit
model (Fig.~\ref{fig,haw-circuit}) focusing on the entanglement
structure.  In our analysis, the size of \textsf{BH} is 6 qubit
($|\textsf{BH}|=2^6, N_\textsf{BH}=6$), \textsf{JR} is 1 qubit
($|\textsf{JR}|=2^1$) and \textsf{ER} is 6 qubit
($|\textsf{ER}|=2^6$).

\subsection{Page curve}
\begin{figure}[H]
   \centering
  \includegraphics[width = 0.48\hsize,clip]{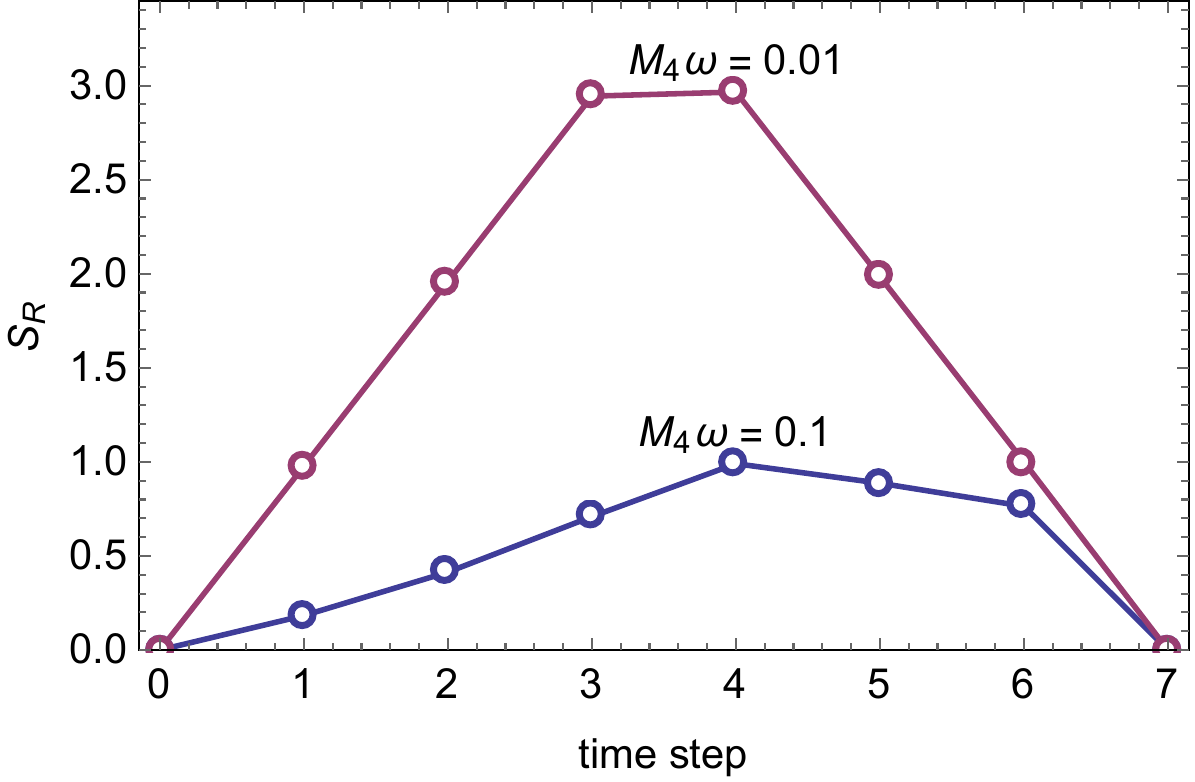}
  \caption{ The Page curve of our circuit model. }
   \label{fig:page_result}
 \end{figure}

As in Sec \ref{evaporation}, we check the amount of information
transferred from the \textsf{BH} to the emitted particles and
investigate whether the black hole in our model evaporates.  For this
purpose, we evaluate the entanglement entropy between \textsf{BH} and
\textsf{R}=\textsf{JR}\,$\cup$\,\textsf{ER}. The evolution of this
quantity provides the Page curve, which is an indicator of the
information transfer from \textsf{BH} to radiation. The result is
shown for $ M_4\, \omega = 0.1$ and $0.01$ in
Fig.~\ref{fig:page_result}.  Here, $ M_4 $ denotes the mass of
the black hole at step 4.
 For $ M_4 \omega=0.01 $ (small $M_4 \omega$), the Page curve has a
 symmetric shape with respect to the middle point of the whole time
 steps. On the other hand, for $ M_4 \omega=0.1$ (large $M_4\omega$),
 the entanglement entropy between \textsf{BH} and \textsf{R} is
 smaller compared to the small $M_4\omega$ case during whole period of
 the evaporation process. For large $M_4\omega$, the black hole radiates
 Hawking particles with weak entanglement (see
 Eq.~\eqref{eq:haw}). Thus the evaporation process is not random
 enough as Page assumed. However, our model results in Page curves for
 both values of $M_4\omega$ because the degree of freedom of \textsf{BH}
 decreases as the time step proceeds due to the action of the SWAP
 gate. The entanglement entropy becomes maximum at step 4 and this
 time step is the Page time of our model.
\subsection{Expectation value of particle number of JR}\label{number_ev}
\anno{To examine the evaporation rate of our model, we calculate the
expectation value of the emitted particle number. In
our model, the Hawking radiation is expressed by the state
(\ref{eq:haw}) and this expression results in the expectatin value of
the  particle
number
\begin{align}
  \ev{n} = \frac{1}{e^{8\pi M\omega}+1}.   \label{eq:expect-a}
\end{align}
Although this distribution is  equivalent to that of the Hawking radiation with
the fermion field~\cite{Hawking_particle_creation}, we are treating
the bosonic particles and the fermionic distribution is result of our
assumption for the state of the Hawking radiation.
}
For the state
$\ket{\psi} = \sum_{ijk} C_{ijk}
\ket{i}_\textsf{BH}\ket{j}_\textsf{JR}\ket{k}_\textsf{ER}$,
the expectation value of the particle number of \textsf{JR}
is  expressed as
\begin{align}
   \ev{n_\textsf{JR}} = \sum_{ik} |C_{i1k}|^2. \label{eq:expect-b}
\end{align}

\begin{figure}[H] 
    \centering
\includegraphics[width = 0.5\hsize,clip]{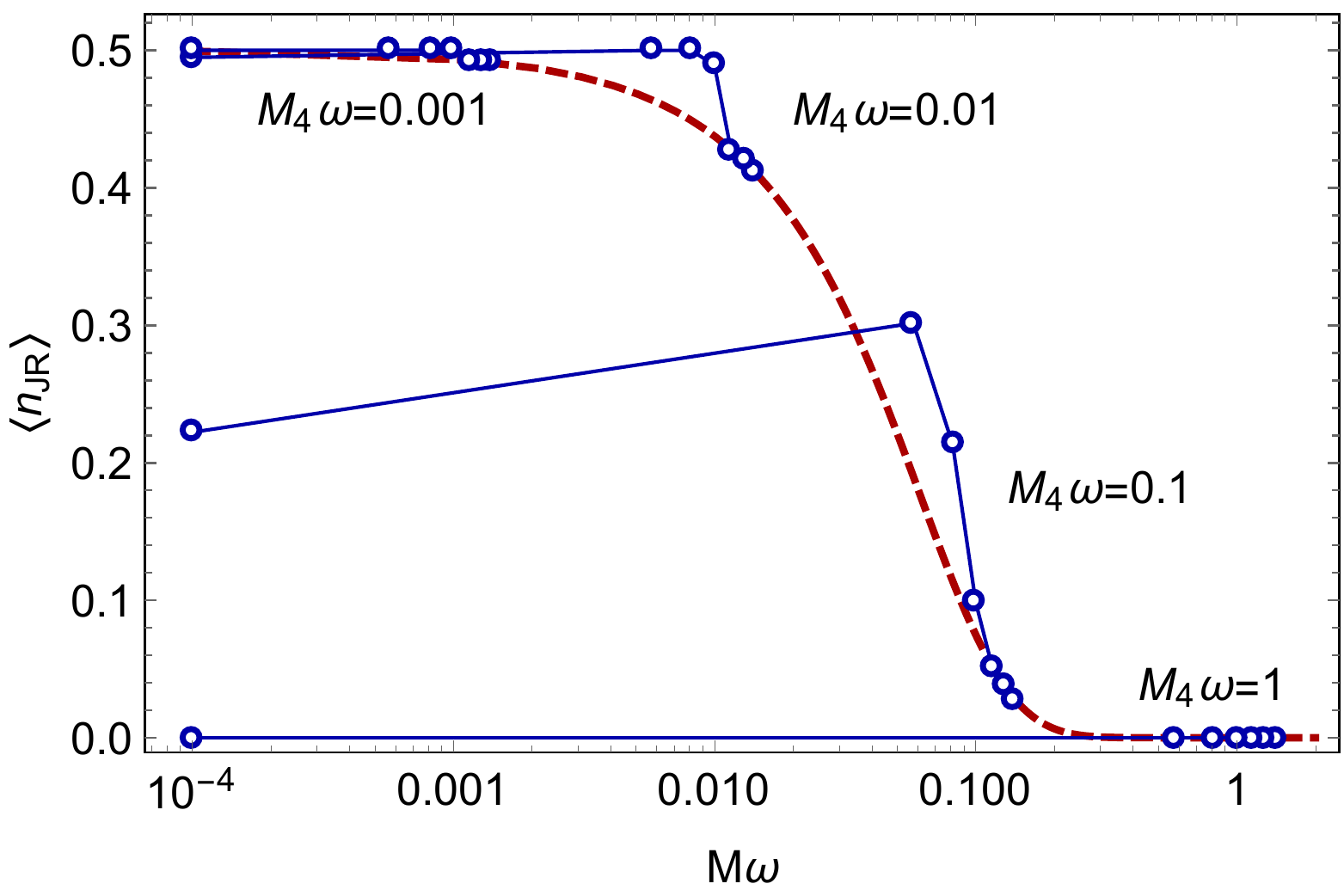}
\caption{The expectation value of the particle number of \textsf{JR}
  as a function of $M\omega$. We plot the particle number from step 1 to
  step 7. The red dotted line corresponds to
  Eq.~\eqref{eq:expect-a}. We assign the value $M\omega=10^{-4}$ to 
    step 7 to represent the particle number at  $M\omega=0$ (the last
    step)  in this plot.} 
  \label{fig:JR_num} 
\end{figure}

We checked these formulas for $ M_4 \omega =1,~0.1,~0.01,~0.001$
(Fig.~\ref{fig:JR_num}). As the black hole evaporates, the mass
 decreases and the time step advances
from right to left.  In our model, $\expval{n_\textsf{JR}}$ well
coincides with the formula \eqref{eq:expect-a} except at the last step
with $M_4\omega\geq 0.1$.  For $M_4\omega\geq
  0.1$, $\expval{n_\textsf{JR}}<0.3$ and this means the emitted pair of
    Hawking particles is weakly entangled. On the other hand, for
    $M_4\omega\leq0.01$, $\expval{n_\textsf{JR}}\approx 0.5$ and this
    means the Hawking pair is nearly maximally entangled.
\noindent
At the last step, the both formula coincides for
$ M_4 \omega\leq 0.01 $ but differs much for $ M_4 \omega\geq 0.1 $.
This disagreement is caused by the different treatment of the Hawking
radiation at the last step in our circuit model.  In the
evaporation process, $M\rightarrow 0$ is realized at the
last step of evaporation. On the other hand, in our model, there is no
CNOT-U gate at the last step (see Fig.~\ref{fig,haw-circuit}), so
sufficient radiation will not be created at the last step. This is the
reason why  large disagreement appeared at the last step for
$M_4\omega\geq 0.1$. However, if the number of qubits is increased and
make the interval of the time step sufficiently small, this
difference is expected to be reduced.
\subsection{Entanglement structure}\label{entstructre}
Let us analyze the entanglement measure in our model.  As we have
already introduced the entanglement entropy as the measure of
entanglement in Sec.~\ref{evaporation}, we introduce here other two
measures needed for our analysis.  The mutual information of the
bipartite system \anno{$\rho_\text{AB}$} is defined
as
\begin{align}
    I(\text{A}\!:\!\text{B})= S_\text{A} + S_\text{B} -
  S_{\text{A}\cup\text{B}}.
\label{mutual-info}
\end{align}
This quantity evaluates how the system is close to the product
state. If the mutual information is zero, the state is the product
state without classical and quantum correlations. The
negativity~\cite{negativity_Vidal,negativity_He} is defined by
  \begin{align}
    \mathcal{N}(\text{A}\!:\!\text{B})= \frac{1}{2}\left(\sum_i|\lambda_i^T|-1\right),\quad\sum_i\lambda_i^T=1,\label{def:negativity}
  \end{align}
  where $\lambda_i^T$ is the eigenvalue of the partially transposed
  state $\rho^{T_\text{A}}_\text{AB}$. The partial transposition
  $T_\text{A}$ for the state
  $\rho = \sum C_{ab:a'b'}\ket{a}\bra{a'}\otimes \ket{b}\bra{b'}$ is
  defined by
  $\rho^{T_\text{A}}:= \sum C_{ab:a'b'}\ket{a'}\bra{a}\otimes
  \ket{b}\bra{b'}$.
  \anno{For a separable bipartite state which has only classical
    correlations, $\lambda_i^T>0$ owing to Peres's partial transpose
    criterion~\cite{Peres1996} and we have $\mathcal{N}=0$. Thus
    $\mathcal{N}>0$ means the state is non-separable (entangled). The
  negativity is an entanglement monotone function, and can be used to
  quantify the entanglement between A and B even if the combined
  system A$\cup$B is not pure. $\mathcal{N}>0$ is the necesary and
  sufficient non-separable condition for $2\otimes2$ and $2\otimes 3$
  bipartite quantum systems. }

\begin{figure}[H] 
    \centering
\includegraphics[width=0.45\linewidth,clip]{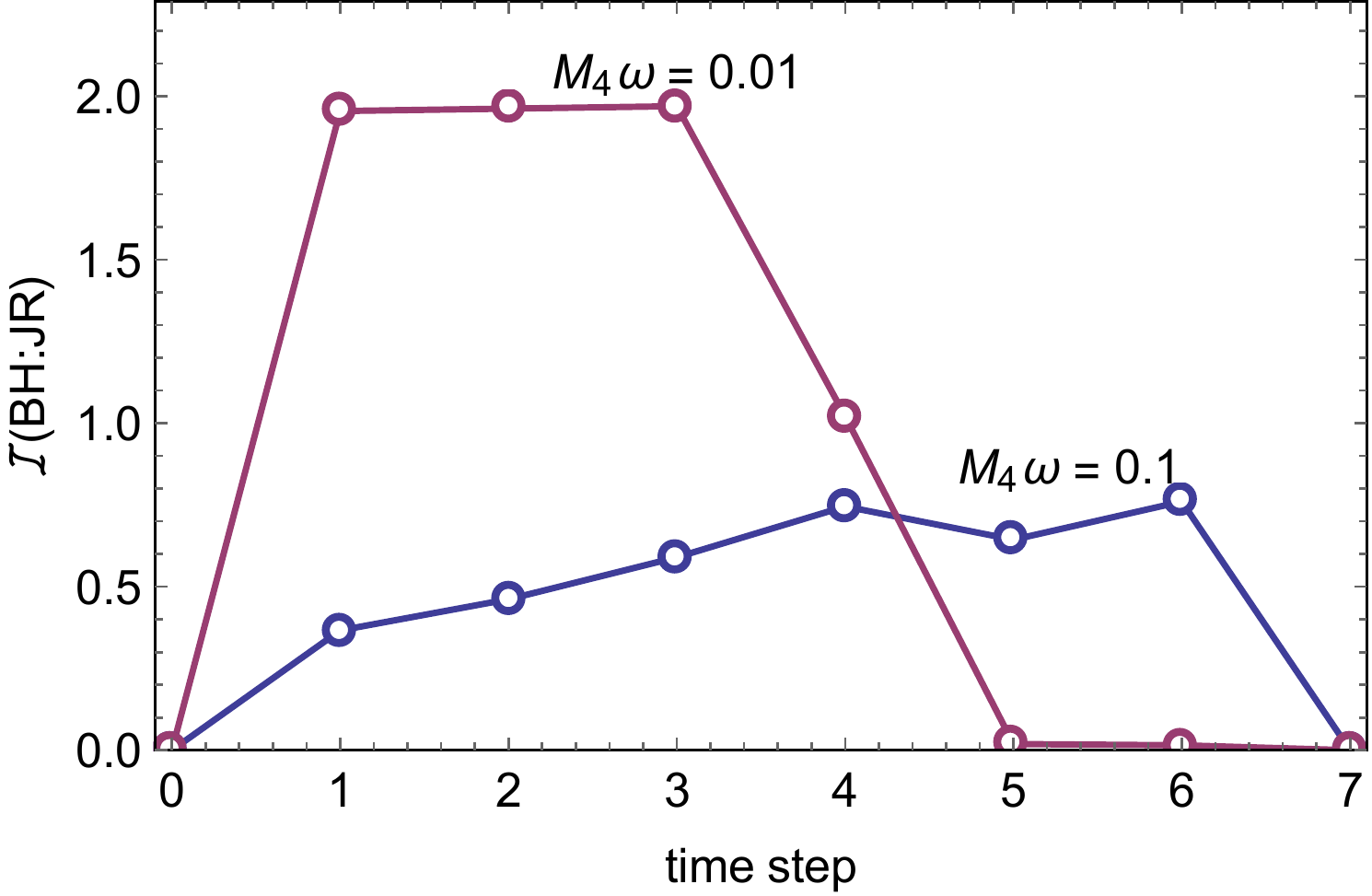}\hspace{0.5cm}
\includegraphics[width=0.45\linewidth,clip]{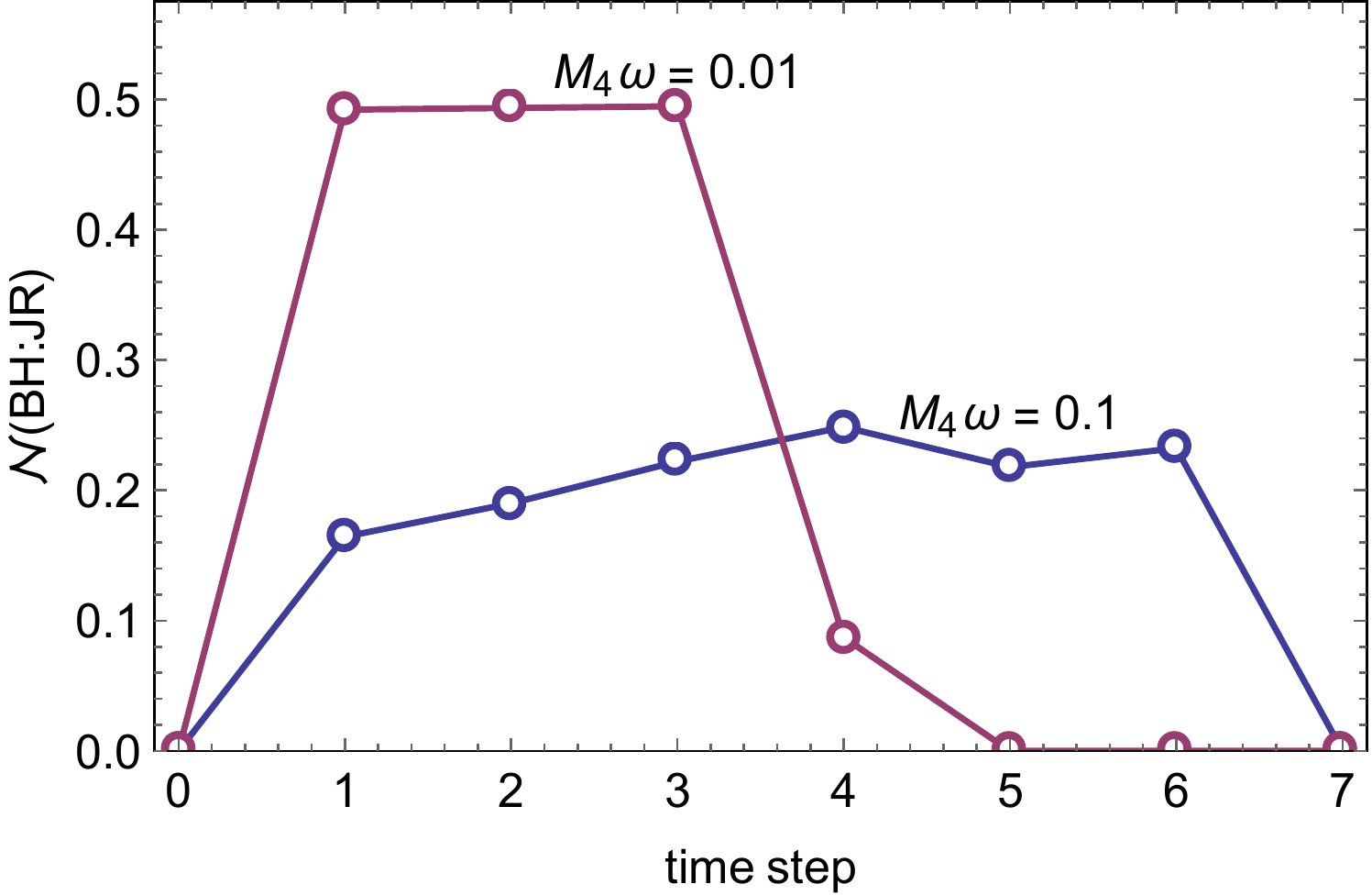}\\
\vspace*{0.5cm}
\includegraphics[width=0.45\linewidth,clip]{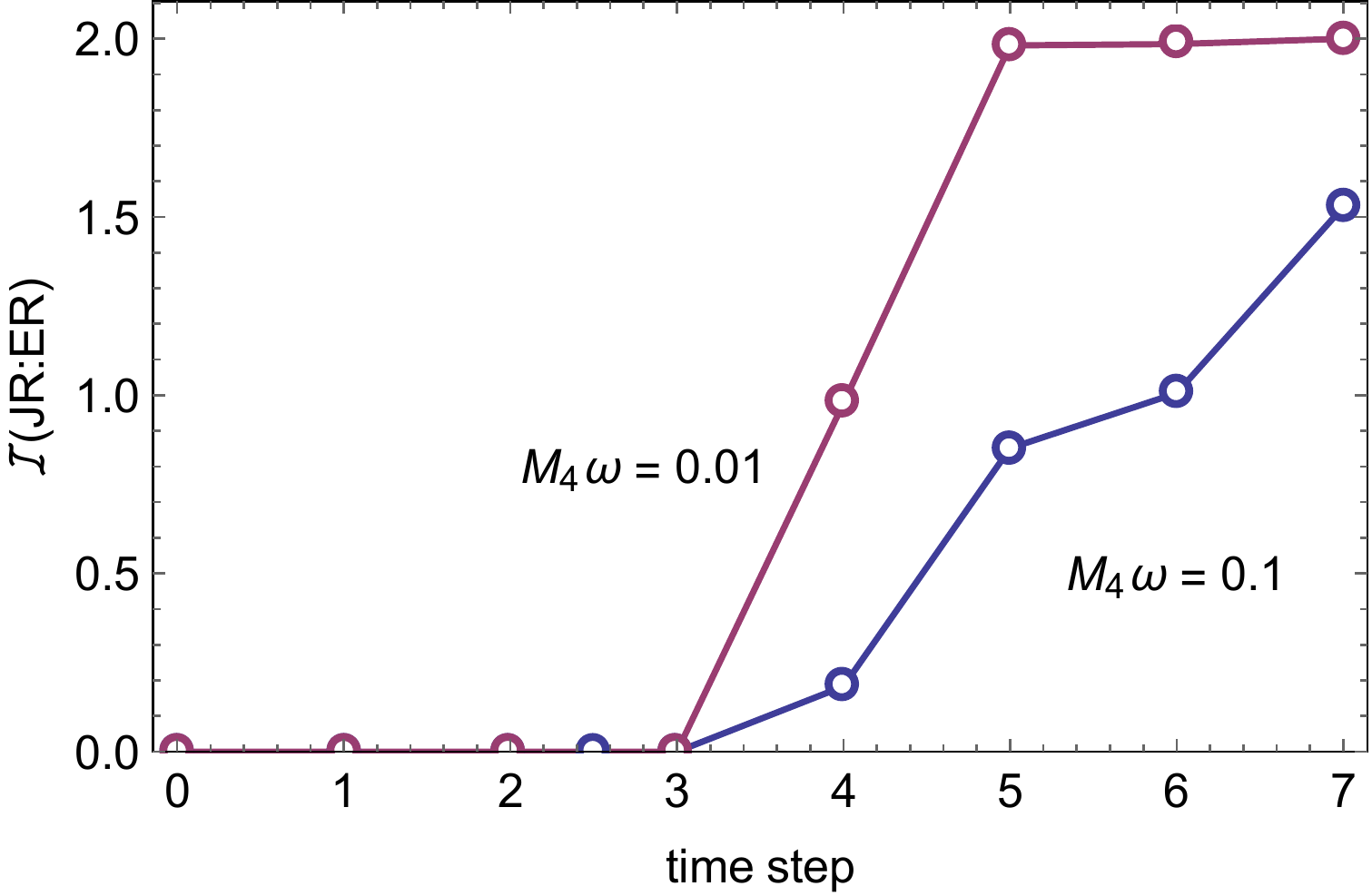}\hspace{0.5cm}
\includegraphics[width=0.45\linewidth,clip]{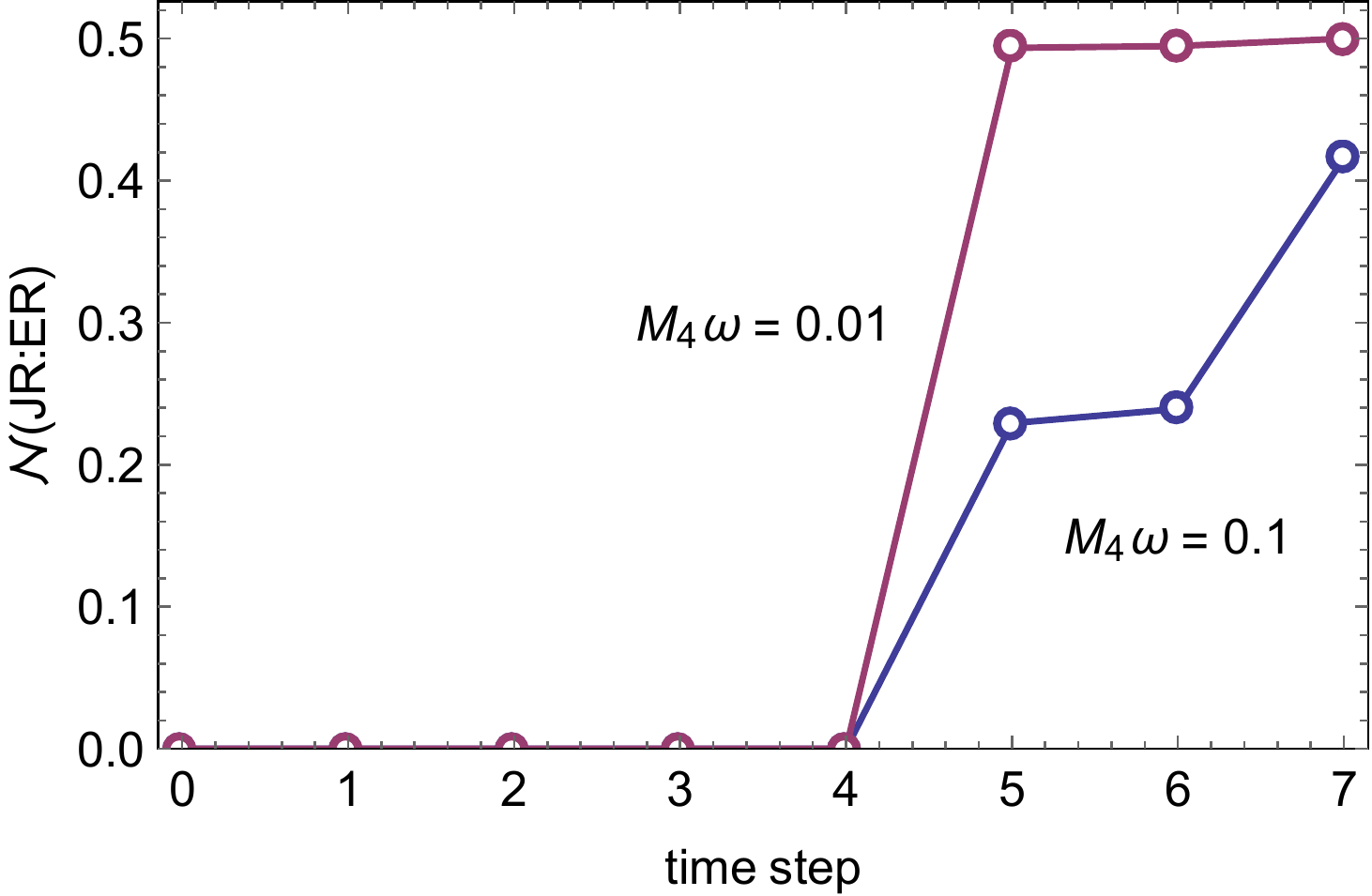}\\
\vspace*{0.5cm}
\includegraphics[width=0.45\linewidth,clip]{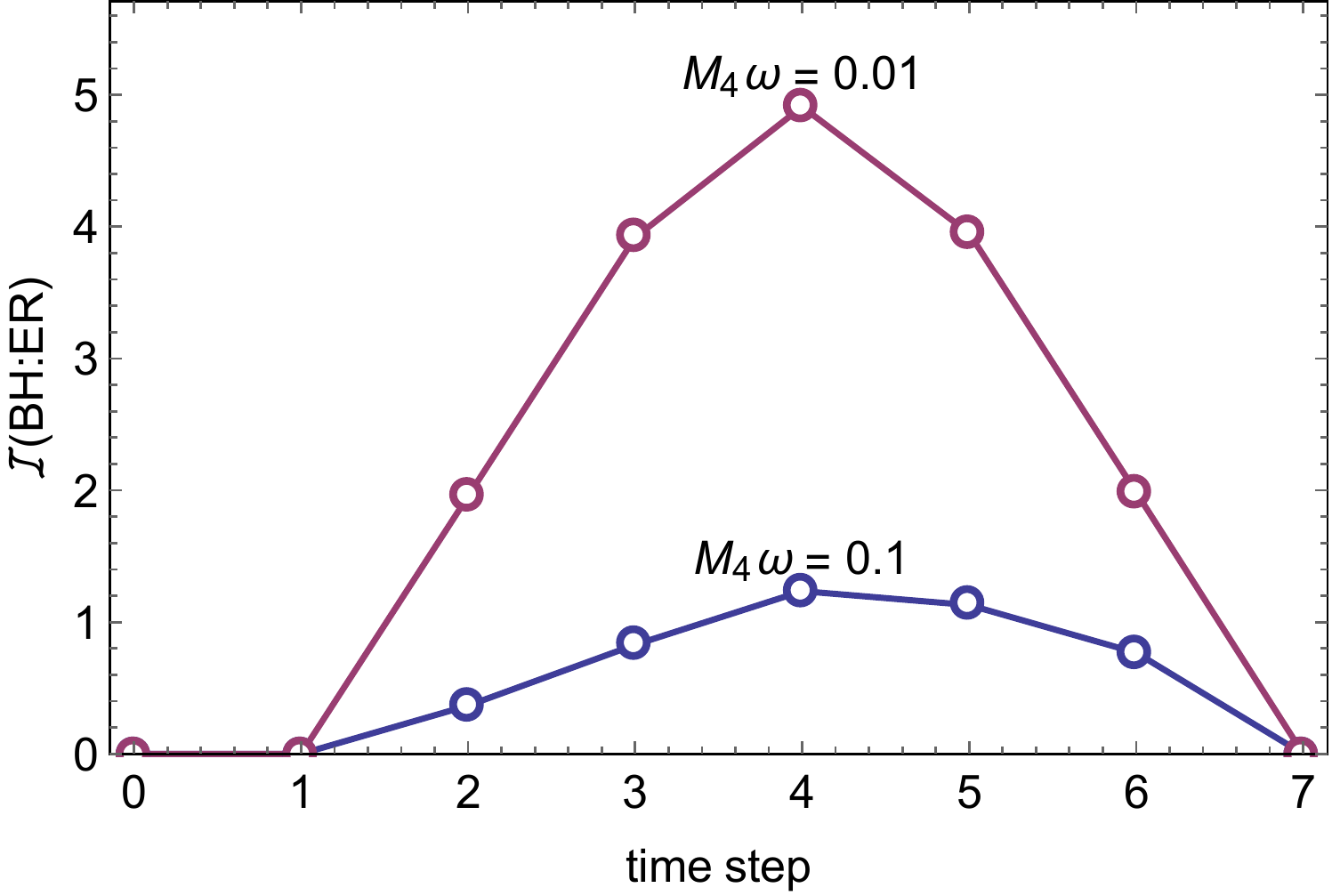}\hspace{0.5cm}
\includegraphics[width=0.45\linewidth,clip]{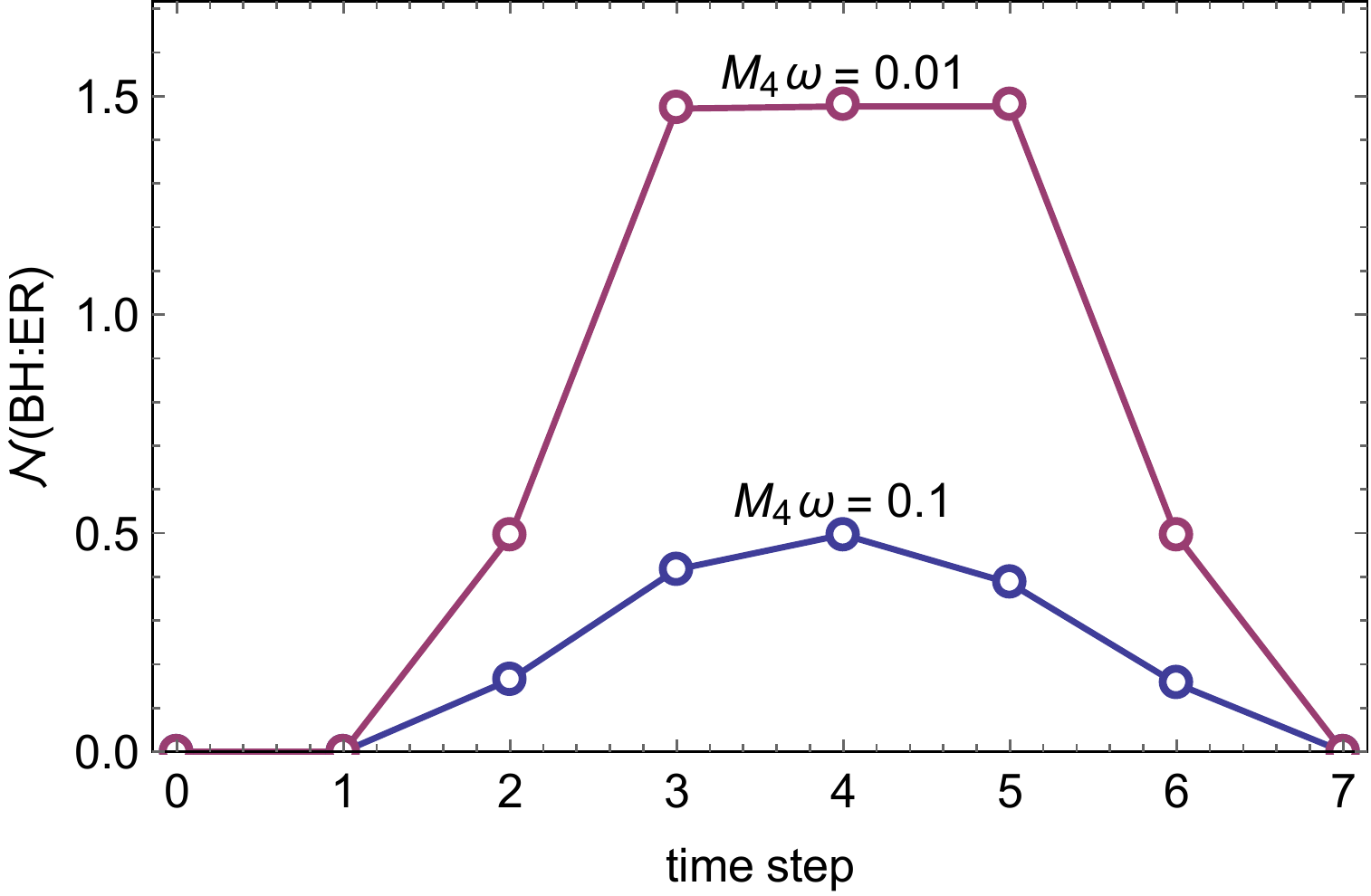}
\caption{Evolution of the mutual information and the negativity.  Upper
  panels: Evolution of the mutual information
  $I(\textsf{BH}\!:\!\textsf{JR})$ and the negativity
  $\mathcal{N}(\textsf{BH}\!:\!\textsf{JR})$. For $M_4\omega=0.01$,
  after step 5, the negativity becomes zero and the mutual information
  has very small but non-zero values. This implies the firewall-like
  structure appears after the Page time. Middle panels: Information
  of emitted radiation, which show how much information is extracted
  from \textsf{BH} at each step. Bottom panels: Entanglement 
  of $\textsf{BH}\cup\textsf{ER}$ subsystem shows nearly the same
   behavior as the Page curve.  }
    \label{fig:result}
  \end{figure}

  We calculate these entanglement measures between \textsf{BH} and
  \textsf{JR}, \textsf{JR} and \textsf{ER}, \textsf{ER} and \textsf{BH} for
  $ M_4\, \omega = 0.1,~0.01 $.  The result is shown in
  Fig.~\ref{fig:result}.
  For $ M_4 \omega=0.01 $, the negativity between \textsf{BH} and
  \textsf{JR} becomes zero at step 5 after the Page time and
  \textsf{BH} and \textsf{JR} are separable. In other words, for the
  Hawking radiation with sufficiently low frequencies, a structure
  similar to the firewall appears between \textsf{BH} and \textsf{JR}.
\anno{However, contrary to the original proposed firewall by AMPS, the
classical correlations remains in our firewall-like structure and we
do not expect high energy phenomena associated to it.}
  On the other hand, for $ M_4 \omega=0.1 $, the values of mutual
  information and negativity between \textsf{BH} and \textsf{JR}
  remain nonzero, and \textsf{BH} and \textsf{JR} are entangled until
  the end of the evaporation.  For sufficiently small $ M_4 \omega $,
  nearly maximally entangled pair of the Hawking particles is created
  by the CNOT-U gate and the qubit strongly entangled with other qubit
  goes through the CNOT-U gate again after the Page time.  However,
  due to the limited amount of entanglement between qubits, which is
  known as the monogamy property of the entanglement, it is impossible
  to add more entanglement and no entangled pairs can be created when
  passing through the second CNOT-U gate after the Page time.  If this
  reasoning for small $ M_4 \omega $ is correct, the emergence of the
  firewall-like structure can be explained.

\section{Analytic Evaluation of State after Page
  time}\label{after-pagetime}
From the results obtained by the numerical calculation, it
turned out that the firewall-like structure appears after the Page
time if $ M_4 \omega $ is sufficiently small.  We intend to explain why
our model shows the firewall-like behavior.
\subsection{Structure of State around Page Time}

According to our numerical calculation, the firewall-like behavior
becomes remarkable as $M_4\omega$ becomes small. To obtain
comprehensive understanding of this structure, we analytically
evaluate the state around the Page time. We introduce the matrix
representation of the U gate and the CNOT gate,
\begin{equation}
  U_n=
  \begin{bmatrix}
    \cos\gamma_n & \sin\gamma_n \\ \sin\gamma_n & -\cos\gamma_n
  \end{bmatrix},\quad
 \mathrm{CNOT}=
 \begin{bmatrix}
    \mathbb{I}_2 & 0 \\ 0 & X
 \end{bmatrix},
\end{equation}
where $\mathbb{I}_2$ is the $2\times 2$ identity matrix, 
$X=\begin{bmatrix} 0 & 1 \\ 1 & 0\end{bmatrix}$ and
\begin{equation}
  \gamma_n=\mathrm{Arctan}\left[\exp\left(-4\pi
      M_4\omega\sqrt{\frac{7-n}{3}}\right)\right],\quad 1\le n\le 7.
\end{equation}
 Then the matrix
representation of the CNOT-U gate is
\begin{align}
  \text{CNOT-U}=\begin{bmatrix} \mathbb{I}_2 & 0 \\ 0 &
    X\end{bmatrix}(\mathrm{U}_n\otimes\mathbb{I}_2) 
  =\begin{bmatrix}
    \cos\gamma_n & 0 & \sin\gamma_n & 0 \\
    0 & \cos\gamma_n & 0 & \sin\gamma_n \\
    0 & \sin\gamma_n & 0 & -\cos\gamma_n \\
    \sin\gamma_n & 0 & -\cos\gamma_n & 0
    \end{bmatrix}.
\end{align}
We prepare $\ket{\text{init}} =
\ket{000000}_\textsf{BH}\ket{0}_\textsf{JR}\ket{000000}_\textsf{ER}$
as the initial state and consider the action of the quantum circuit on
this state.

\begin{figure}[H]
    \centering
    \includegraphics[width = 0.55\hsize,clip]{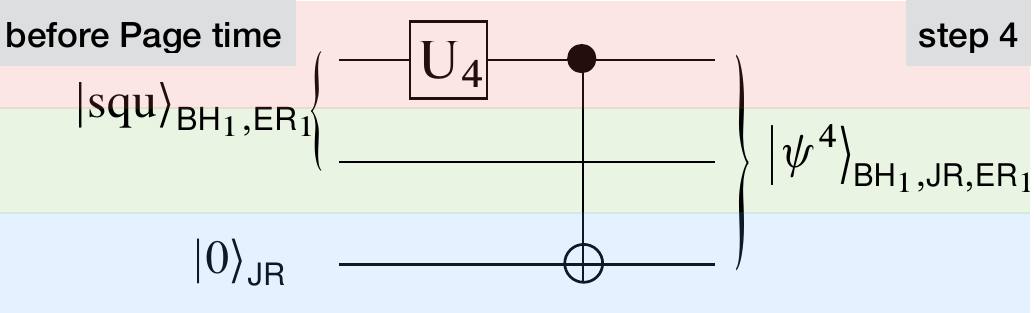}
       \caption{Quantum circuit from before the Page time to step 4.}
    \label{fig:just4}
 \end{figure}

\noindent
\textbf{Before the Page time:} The Page time corresponds to step 4 and
the circuit up to just before step 4 is equivalent to the following
circuit diagram (Fig.~\ref{fig,before4}). We rearranged order of qubit
and changed the number of labels from those defined in the original
circuit in Fig.~\ref{fig,haw-circuit}.
\begin{figure}[H]
    \centering
    \includegraphics[width = 0.45\hsize,clip]{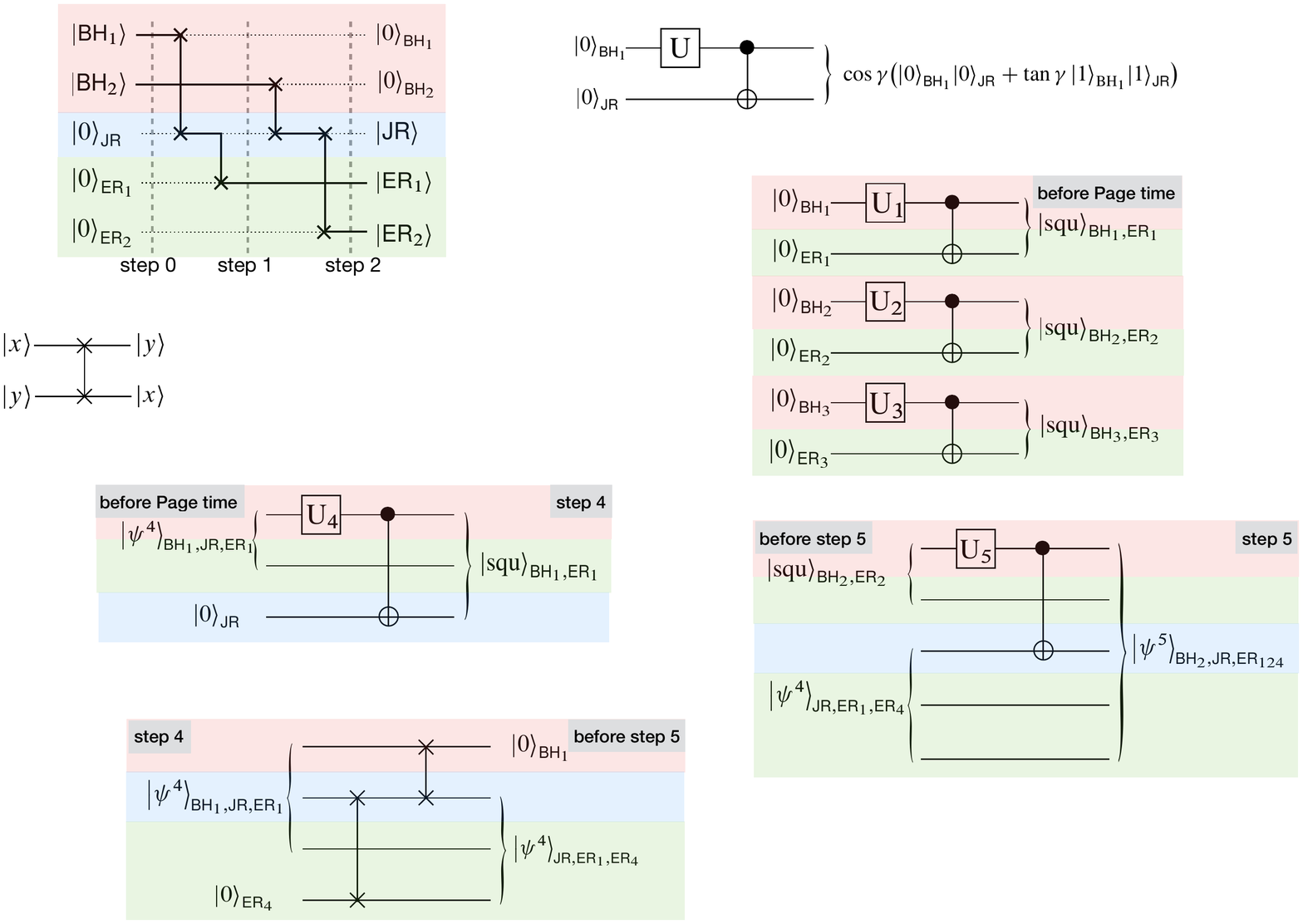}
    \caption{Quantum circuit before the Page time (before step
       4). $\ket{\text{squ}} := \cos \gamma \ket{00}+\sin\gamma \ket{11}$.}
    \label{fig,before4}
 \end{figure}
\noindent
The state of the total system is
\begin{align}
  \ket{\psi^{\text{before step 4}}} = \ket{\text{squ}}_{\textsf{BH}_1,\textsf{ER}_1}\ket{\text{squ}}_{\textsf{BH}_2,\textsf{ER}_2} \ket{\text{squ}}_{\textsf{BH}_3,\textsf{ER}_3} \ket{0}_{\textsf{BH}_{456},\textsf{JR},\textsf{ER}_{456}}.
\end{align}
\noindent
\textbf{At step 4: } The circuit just before the Page time to step 4
(the Page time) can be drawn as Fig.~\ref{fig:just4}. Only
$\textsf{BH}_1$, $\textsf{JR}$ and $\textsf{ER}_1$ get involved.
\noindent
The state at step 4 is obtained as
\begin{align}
 \ket{\psi^4}_{\textsf{BH}_1,\textsf{JR},\textsf{ER}_1} &=
                                                          \text{CNOT}_{\textsf{BH}_1,\textsf{JR}}\,
 U_{\textsf{BH}_1}\ket{\text{squ}}_{\textsf{BH}_1,\textsf{ER}_1}\ket{0}_{\textsf{JR}}\notag\\
  &=
    \text{CNOT}_{\textsf{BH}_1,\textsf{JR}}\biggl(\cos\gamma_1\Bigl[\cos\gamma_4\ket{0}+\sin\gamma_4\ket{1}\Bigr]_{\textsf{BH}_1}\ket{0}_{\textsf{ER}_1}
    \notag \\
  &\qquad\qquad\qquad\qquad +\sin\gamma_1\Bigl[\sin\gamma_4\ket{0}+\cos\gamma_4\ket{1}\Bigr]_{\textsf{BH}_1}\ket{1}_{\textsf{ER}_1}
    \biggr)\ket{0}_\textsf{JR} \notag
\\
 &=
   \cos\gamma_1\Bigl[\cos\gamma_4\ket{000}+\sin\gamma_4\ket{101}\Bigr]_{\textsf{BH}_1,
   \textsf{JR}, \textsf{ER}_1} \notag \\
  &\qquad
    +\sin\gamma_1\Bigl[\sin\gamma_4\ket{010}-\cos\gamma_4\ket{111}\Bigr]_{\textsf{BH}_1,
    \textsf{JR}, \textsf{ER}_1}.
 \label{step4}
\end{align}
In the low frequency limit $M_4\omega\rightarrow 0$, squeezing
  parameters are  $\gamma_1,\gamma_4\rightarrow \pi/4$ and
\begin{equation}
  \ket{\psi^4}_{\textsf{BH}_1, \textsf{JR},
    \textsf{ER}_1}\longrightarrow
  H_\textsf{JR}\frac{1}{\sqrt{2}}\Bigl[\ket{000}+\ket{111}\Bigr]_{\textsf{BH}_1,
  \textsf{JR}, \textsf{ER}_1},
\end{equation}
where 
  $H_\textsf{JR}:=\begin{bmatrix} 1 & 1 \\ 1 & -1 \end{bmatrix}$ acts
  on the qubit of \textsf{JR}.  The appeared state
$\qty(\ket{000}+\ket{111})/\sqrt{2}$ is the GHZ state, which is the
maximally entangled triqubit state. If one qubit in the GHZ state is
traced out, the remaining subsystem with two qubits becomes
separable. Since an unitary operator acting only on a single qubit
does not affect the entanglement structure, \textsf{BH} and
\textsf{JR} become separable at step 4 in the low frequency
  limit if \textsf{ER} is traced out. As we will see, this leads to the
firewall-like structure at step 5.

\begin{figure}[H]
    \centering  
     \includegraphics[width = 0.6\hsize,clip]{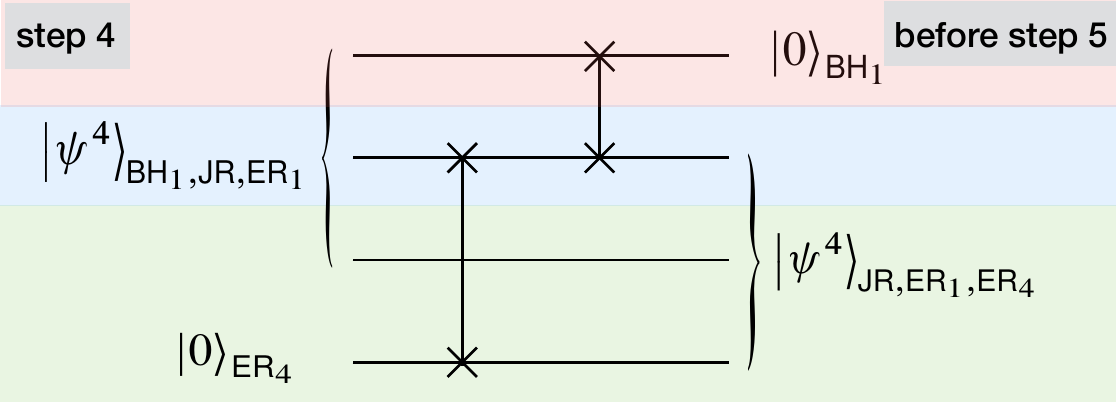}
    \caption{Quantum circuit for step 4 to step $4'$ = before step 5.}
    \label{fig:before5}
 \end{figure}

\noindent
\textbf{At before step 5:}
The circuit diagram is Fig.~\ref{fig:before5}.
Qubits pass through the SWAP gate  from step 4 to step 5.
\noindent
At step 4, one qubit in \textsf{BH} and two qubits in \textsf{JR} and
\textsf{ER} are entangled. At just before step 5, one qubit in
\textsf{BH} is $\ket{0}$ state, and one qubit in \textsf{JR} and two
qubit in \textsf{ER} are entangled.

\begin{figure}[H] 
    \centering
    \includegraphics[width = 0.55\hsize,clip]{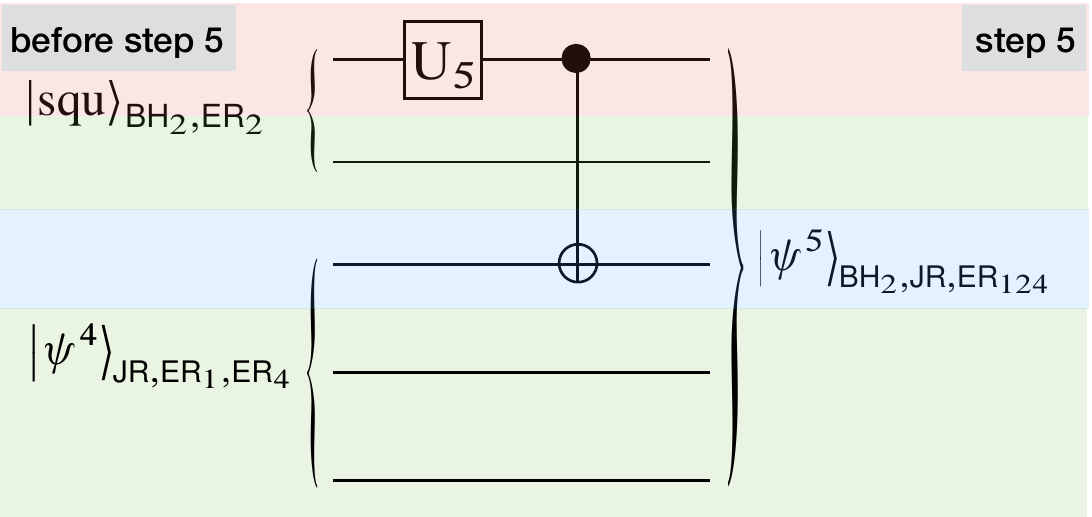}
    \caption{Quantum circuit for step $4'$ to step 5.}
    \label{fig:just5} 
 \end{figure}

\noindent
\textbf{At step 5:}
The quantum circuit up to step 5 is shown in Fig.~\ref{fig:just5}.
\noindent
The state at step 5 is obtained as
\begin{align}
  \ket{\psi^5}_{\textsf{BH}_{12},\textsf{JR},\textsf{ER}_{124}} = \ket{0}_{\textsf{BH}_1}\ket{\psi^5}_{\textsf{BH}_2,\textsf{JR},\textsf{ER}_{124}},
\end{align}
where
\begin{align}
 \ket{\psi^5}_{\textsf{BH}_2,\textsf{JR},\textsf{ER}_{124}}
&=\mathrm{CNOT}_{\textsf{BH}_2,
  \textsf{JR}}\,U_{\textsf{BH}_2}\ket{\text{squ}}_{\textsf{BH}_2,
  \textsf{ER}_2}\ket{\psi^4}_{\textsf{JR}, \textsf{ER}_{14}} \notag \\
&=\mathrm{CNOT}_{\textsf{BH}_2,
  \textsf{JR}}\biggl[\cos\gamma_2\left(\cos\gamma_5\ket{00}+\sin\gamma_5\ket{10}\right)
  \notag \\
&\quad\qquad\qquad\qquad
  +\sin\gamma_2\left(\sin\gamma_5\ket{01}-\cos\gamma_5\ket{11}\right)\biggr]_{\textsf{BH}_2,
  \textsf{ER}_2}\ket{\psi^4}_{\textsf{JR}, \textsf{ER}_{14}} \notag \\
%
%
%
 &=\cos\gamma_1\cos\gamma_2\cos\gamma_4\cos\gamma_5\biggl(
 \ket{0}_{\textsf{BH}_2}\Bigl[\ket{0}+\tan\gamma_2\tan\gamma_5\ket{1}\Bigr]_{\textsf{ER}_2}
   \notag \\
&\qquad\times\Bigl[\ket{000}+\tan\gamma_4\ket{101}+\tan\gamma_1\tan\gamma_4\ket{010}-\tan\gamma_1\ket{111}\Bigr]_{\textsf{JR},
  \textsf{ER}_{14}} \notag \\
&\qquad
  +\ket{1}_{\textsf{BH}_2}\Bigl[\tan\gamma_5\ket{0}-\tan\gamma_5\ket{1}\Bigr]_{\textsf{ER}_2}
  \notag \\
&\qquad\times
  \Bigl[\ket{100}+\tan\gamma_4\ket{001}+\tan\gamma_1\gamma_4\ket{110}-\tan\gamma_1\ket{011}\Bigr]_{\textsf{JR}, \textsf{ER}_{14}}
\biggr).
\end{align}

\subsection{Negativity and Mutual Information at step 4 and 5}
As we have obtain the state at step 4 and step 5, it is possible to
investigate the entanglement structure by evaluating the mutual
information and the negativity  between
\textsf{BH} and \textsf{JR}.

\subsubsection{At step 4}
The reduced density matrix $\rho^4_{\textsf{BH}_1\cup\textsf{JR}}$ at
step 4 is 
\begin{align}
  \rho^4_{\textsf{BH}_1\cup\textsf{JR}}&=\Tr_{\textsf{ER}}\left[\ket{\psi^4}\bra{\psi^4}_{\textsf{BH}_1,
                                      \textsf{JR},
                                      \textsf{ER}_1}\right] \notag \\
 &=\left(\cos^2\gamma_1\cos^2\gamma_4+\sin^2\gamma_1\sin^2\gamma_4\right)
   \ket{00}\bra{00}+\frac{1}{2}\cos 2\gamma_1\sin
   2\gamma_4\ket{00}\bra{11} \notag \\
  &\quad +\frac{1}{2}\cos 2\gamma_1\sin
    2\gamma_4\ket{11}\bra{00}+\left(
    \cos^2\gamma_1\sin^2\gamma_4+\sin^2\gamma_1\cos^2\gamma_4\right)\ket{11}\bra{11}
    \notag \\
  &=
    \frac{1}{2}\begin{bmatrix}
      1+\cos2\gamma_1\cos2\gamma_4 & 0 & 0 &
      \cos 2\gamma_1\sin 2\gamma_4 \\
      0 & 0 & 0 & 0 \\
      0 & 0 & 0 & 0 \\
      \cos 2\gamma_1\sin 2\gamma_4 & 0 & 0 &
      1-\cos 2\gamma_1\cos 2\gamma_4
    \end{bmatrix}.
\end{align}

\begin{figure}[H]
    \centering
    \includegraphics[width =
    0.48\hsize,clip]{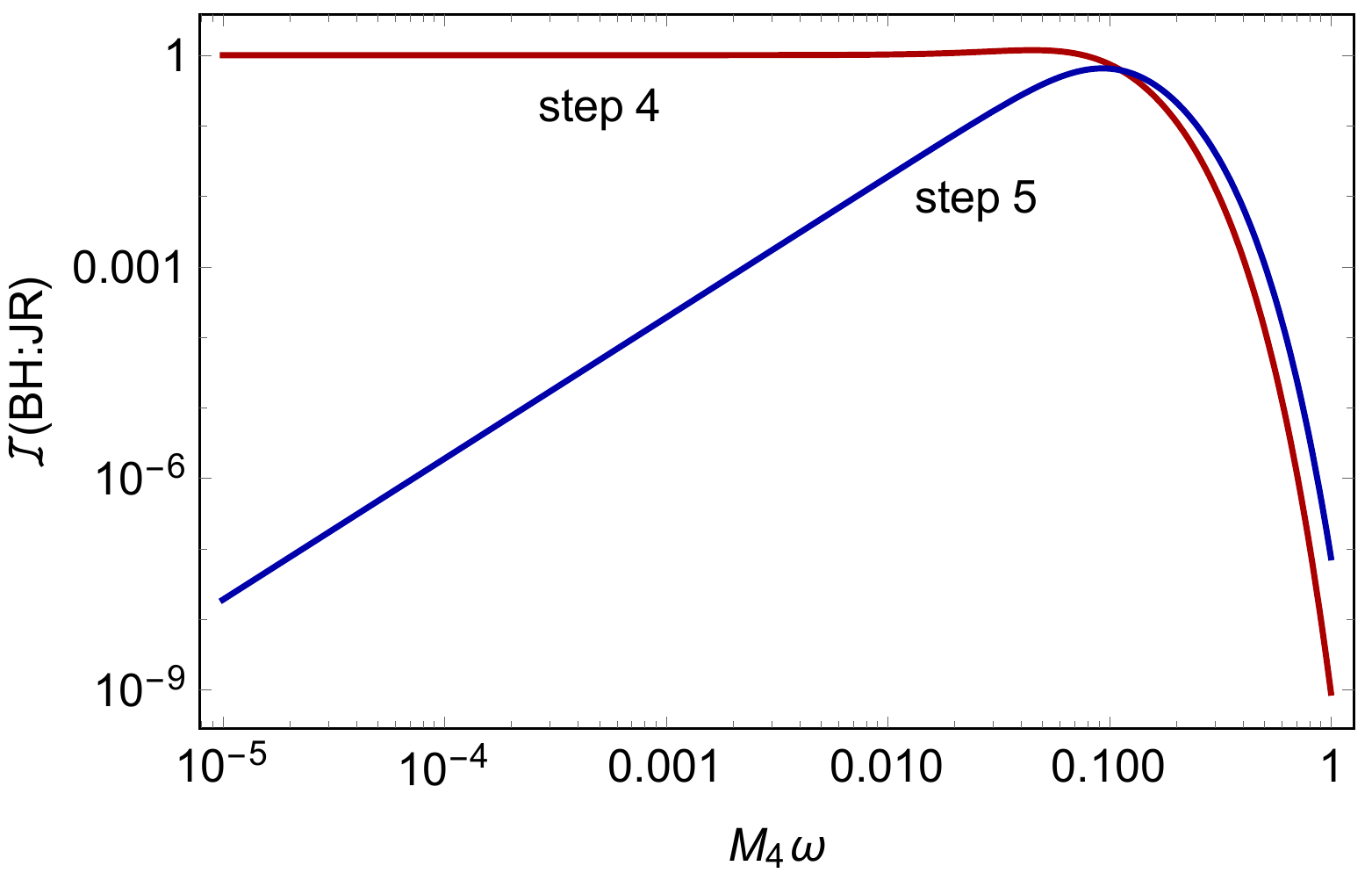}%
    \hspace{0.4cm}
    \includegraphics[width =0.48\hsize,clip]{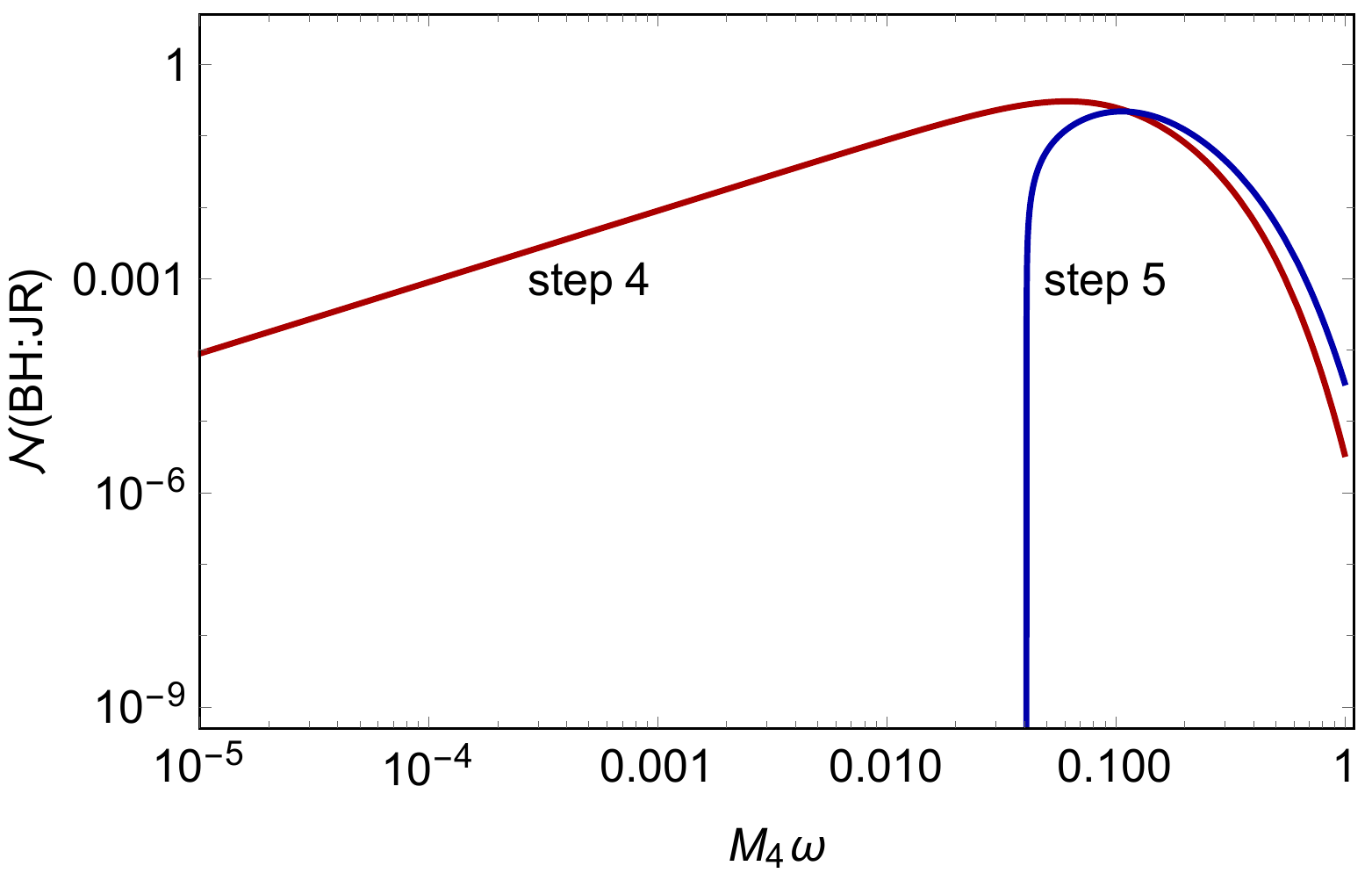}
    \caption{Left panel: $M_4\omega$ dependence of the mutual
      information between $\textsf{BH}_1$ and \textsf{JR} at step 4 and 5.
      Right panel: $M_4\omega$ dependence of the negativity between
      $\textsf{BH}_2$ and \textsf{JR} at step 4 and 5. The
        negativity becomes zero for $M_4\omega\leq 0.041$ at step 5
        and $\textsf{BH}_2\!\cup\!\textsf{JR}$ system becomes separable. }
    \label{fig:step5neg}
 \end{figure}

To obtain the mutual information $I(\textsf{BH}_1\!:\!\textsf{JR})$, we
prepare the reduce states
\begin{align}
  \rho^4_{\textsf{BH}_1}&=\Tr_{\textsf{JR}}\left[\rho^4_{\textsf{BH}_1\cup\textsf{JR}}\right]=\frac{1}{2}\begin{bmatrix}1+\cos 2\gamma_1
    \cos 2\gamma_4 & 0 \\ 0 &
    1-\cos 2\gamma_1
    \cos 2\gamma_4  \end{bmatrix},\notag \\
\rho^4_{\textsf{JR}}&=\Tr_{\textsf{BH}_1}\left[\rho^4_{\textsf{BH}_1\cup\textsf{JR}}\right]=\frac{1}{2}\begin{bmatrix}1+\cos 2\gamma_1
    \cos 2\gamma_4 & 0 \\ 0 &
    1-\cos 2\gamma_1
    \cos 2\gamma_4  \end{bmatrix}.
\end{align}
Eigenvalues of these states are
\begin{align}
  &\lambda_{\textsf{BH}_1\cup\textsf{JR}}=\frac{1}{2}\left(1\pm\cos
    2\gamma_1\right),\quad
    \lambda_{\textsf{BH}_1}=\lambda_{\textsf{JR}}=\frac{1}{2}(1\pm\cos
    2\gamma_1\cos 2\gamma_4),
\end{align}
and the mutual information is 
\begin{equation}
  I(\textsf{BH}_1\!:\!\textsf{JR})=
-\sum_{i=\textsf{BH}_1, \textsf{JR}}\lambda_i\log_2\lambda_i
+\sum_{i=\textsf{BH}_1\cup\textsf{JR}}\lambda_i\log_2\lambda_i.
\end{equation}
The red line in the left panel of Fig.~\ref{fig:step5neg} shows the mutual
information as a function of $M_4\omega$. It behaves as
  \begin{equation}
I\sim
      \begin{cases}
        1+16\pi^2(M_4\omega)^2 &\quad \text{for} \quad M_4\omega\ll 0.1\\ 
        16\sqrt{2}\pi (M_4\omega)e^{-8\pi M_4\omega} &\quad \text{for}
        \quad M_4\omega\gg 0.1
      \end{cases}
  \end{equation}
The negativity is obtained using eigenvalues of the partial
transposed state $\rho^{4T_\textsf{JR}}_{\textsf{BH}\cup\textsf{JR}}$
\begin{equation}
\frac{1}{2}\left(1\pm\cos 2\gamma_1\cos
  2\gamma_4\right),\quad\pm\frac{1}{2}\cos 2\gamma_1\sin 2\gamma_4,
\end{equation}
and 
\begin{equation}
  \mathcal{N}(\textsf{BH}_1\!:\!\textsf{JR})=\frac{1}{2}\cos 2\gamma_1\sin
  2\gamma_4. 
\end{equation}
The red line in the right panel of Fig.~\ref{fig:step5neg} shows the
negativity as a function of $M_4\omega$. The negativity at step 4 has
a peak at $M_4\omega\approx 0.1$ and behaves as
  \begin{equation}
\mathcal{N}\sim
      \begin{cases}
        2\sqrt{2}\pi M_4\omega &\quad \text{for} \quad M_4\omega\ll 0.1\\ 
        e^{-4\pi M_4\omega} &\quad \text{for}
        \quad M_4\omega\gg 0.1
      \end{cases}
  \end{equation}
  As we have mentioned,
  $\ket{\psi^4}_{\textsf{BH}_1, \textsf{JR}, \textsf{ER}_1}$ becomes
  the GHZ state for $M_4\omega\rightarrow 0$. Corresponding to this,
  the state $\rho^4_{\textsf{BH}_1\cup\textsf{JR}}$ becomes separable
  in this limit. The von Neuman entropies for this state are
  $S(\textsf{BH}_1)=S(\textsf{JR})=S(\textsf{BH}_1\!\cup\!\textsf{JR})=1$
  (bit).
  Thus the mutual information has the value 1 (bit), which means
  $\textsf{BH}_1$ and \textsf{JR} have the perfect classical
  correlation.

\subsubsection{At step 5}
The reduced density matrix $\rho^5_{\textsf{BH}_2\cup\textsf{JR}}$
 is
\begin{equation}
  \rho^5_{\textsf{BH}_2\cup
    \textsf{JR}}=\Tr_\textsf{ER}\Bigl[\ket{\psi^5}\bra{\psi^5}_{\textsf{BH}_2,
      \textsf{JR}, \textsf{ER}_{14}}\Bigr]=
\begin{bmatrix}
  CA & 0 & 0 & EA \\
  0 & CB & EB & 0 \\
  0 & EB & DB & 0 \\
  EA & 0 & 0 & DA
\end{bmatrix},
\end{equation}
where
\begin{align*}
  &A=\frac{1}{2}(1+\cos 2\gamma_1\cos 2\gamma_4),\quad
    B=\frac{1}{2}(1-\cos 2\gamma_1\cos 2\gamma_4),\quad
    C=\frac{1}{2}(1+\cos 2\gamma_2\cos 2\gamma_5), \\
  &D=\frac{1}{2}(1-\cos 2\gamma_2\cos 2\gamma_5),\quad
    E=\frac{1}{2}\cos 2\gamma_2\sin 2\gamma_5.
\end{align*}
 To obtain the mutual information between $\textsf{BH}_2$ and
   \textsf{JR},  we  calculate eigenvalues of the states
 $ \rho_\textsf{JR}, \rho_{\textsf{BH}_2}$ and $\rho_{\textsf{BH}_2\cup\textsf{JR}} $.
 Eigenvalues $\lambda_i$ ($i=$ \textsf{JR}, $\textsf{BH}_2$,
 $\textsf{BH}_2\!\cup\!\textsf{JR}$) are
\begin{align}
 &\lambda_\textsf{JR}=\frac{1}{2}(1\pm\cos 2\gamma_1\cos 2\gamma_2\cos
   2\gamma_4\cos 2\gamma_5),\quad
\lambda_\textsf{BH}=\frac{1}{2}(1\pm\cos 2\gamma_2\cos 2\gamma_5),\notag \\
&\lambda_{\textsf{BH}_2\cup\textsf{JR}}=\frac{1}{2}(1\pm\cos 2\gamma_1\cos
  2\gamma_4)\cos^2 \gamma_2,~ \frac{1}{2}(1\pm\cos 2\gamma_1\cos 2\gamma_4)\sin^2
\gamma_2.
\end{align}
The mutual information between $\textsf{BH}_2$ and \textsf{JR} as a
function of $M_4\omega$ is shown in Fig.~\ref{fig:step5neg} (the blue
line in the left
panel). It behaves as $I\approx (40\pi^2/3)(M_4\omega)^2$ for
$ M_4\omega\ll0.1 $ and $I\rightarrow 0$ for
  $M_4\omega\rightarrow 0$ limit.

\begin{figure}[H]
  \centering
  \includegraphics[width = 0.4\hsize,clip]{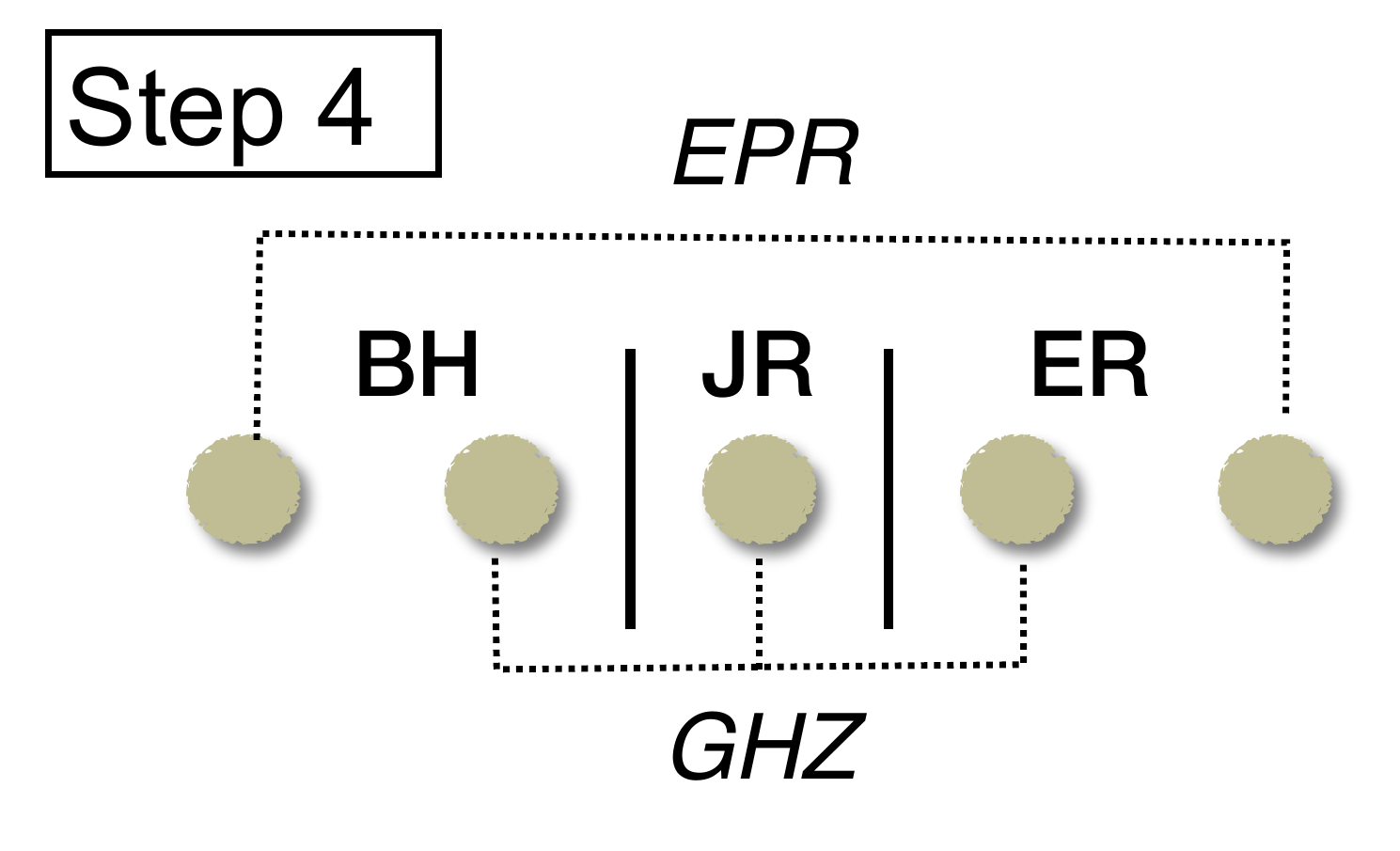}\hspace{1cm}
  \includegraphics[width = 0.4\hsize,clip]{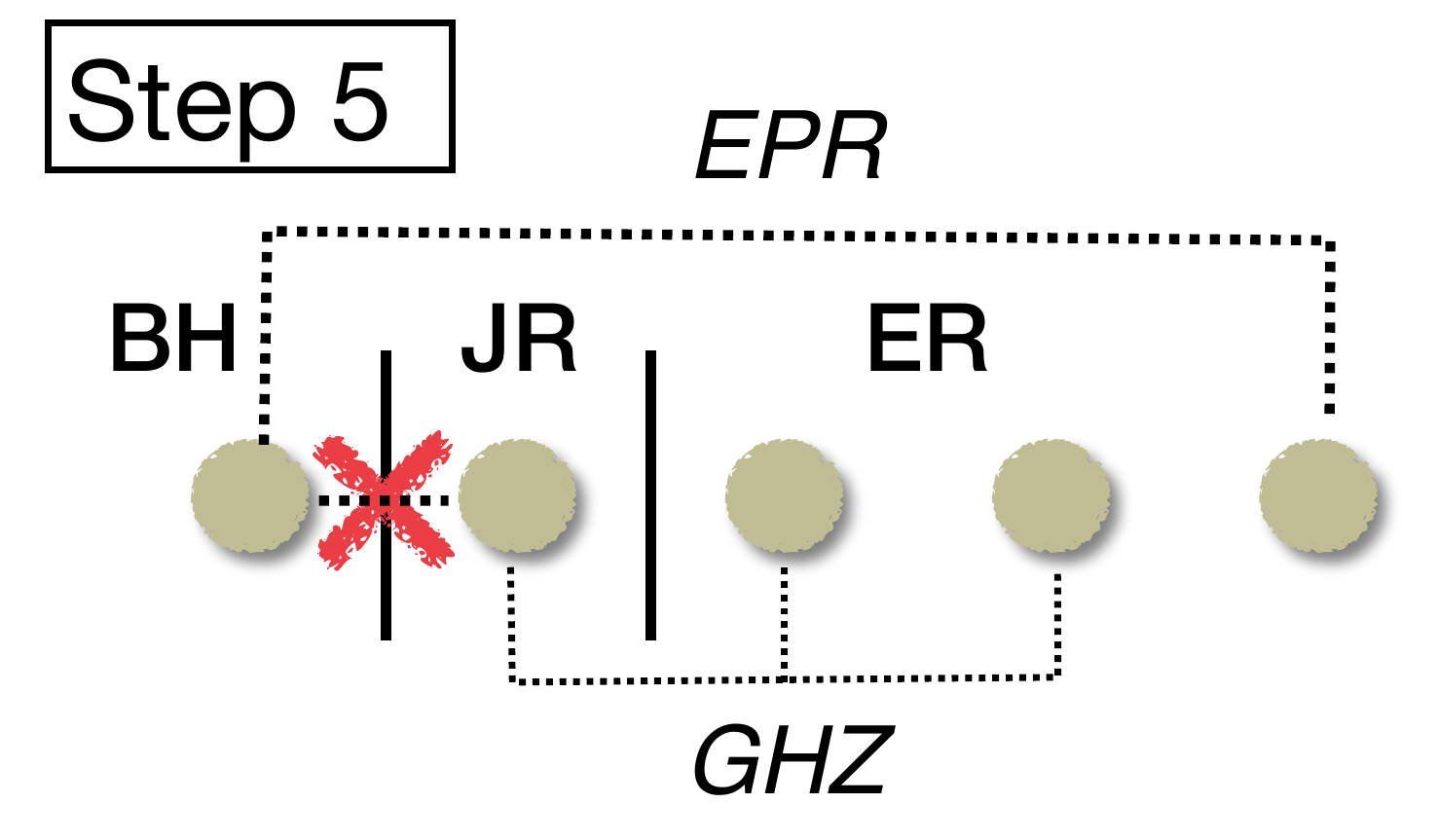}
\caption{The state at step 4 and step 5 (at the Page
  time and next step) in the low frequency limit ($M_4\omega\rightarrow
  0$).}
  \label{fig_result_state}
\end{figure}

To evaluate the negativity, we obtain the eigenvalues of the partially
transposed state $\rho^{T_\textsf{JR}}_{\textsf{BH}_2,\textsf{JR}}$ as
\begin{align}
  \lambda_i&=\frac{(C+D)A\pm\sqrt{(C-D)^2A^2+4E^2B^2}}{2},\quad 
\frac{(C+D)B\pm\sqrt{(C-D)^2B^2+4E^2A^2}}{2},
\end{align}
and the negativity is given by
$\mathcal{N}=\left(\sum_i|\lambda_i|-1\right)/2$.  The result is shown
in the right panel in Fig.~\ref{fig:step5neg} (the blue line). The
negativity becomes exactly zero for $M_4\omega\leq 0.041$. Thus, at
step 5 (next step to the Page time), our model shows $\textsf{BH}_2$ and
\textsf{JR} become separable for low frequency modes satisfying
$M_4\omega < 0.041$ and have only the classical
  correlation. This separable state corresponds to the firewall. In
  the $M_4\omega\rightarrow 0$ limit ($\gamma\rightarrow\pi/4$),
the reduced density matrix at step 5 is
\begin{align}
  \rho_{\textsf{BH}_2\cup\textsf{JR}} = &\frac{1}{4}\qty(\op{0}+\op{1})_\textsf{JR}\otimes \qty(\op{0}+\op{1})_{\textsf{BH}_2}=\frac{1}{4}\,\mathbb{I}_{4}.
\end{align}
Thus, $\textsf{BH}_2$ and \textsf{JR} are in a product state with
  no classical correlation (random). For $0<M_4\omega\ll 0.041$, the
  state is separable with weak classical correlations.  The schematic
diagram representing the quantum states at step 4 and step 5 is shown
in Fig.~\ref{fig_result_state}.
\noindent
Three qubits in \textsf{BH}, \textsf{JR} and \textsf{ER} form the GHZ
state at step 4, which is nearly maximally entangled. At step 5, the CNOT-U
gate acts on \textsf{BH} and \textsf{JR}. However, as \textsf{BH} and
\textsf{JR} are already maximally entangled with other qubits, and no
entanglement can be shared between \textsf{BH} and \textsf{JR} by the
monogamy property of the multipartite entanglement.  This is
the reason why the entanglement between \textsf{BH} and \textsf{JR} is
lost and the firewall-like structure arises for
$M_4\omega\leq 0.041$.

\section{Summary and Discussion}
We constructed a quantum circuit model that realizes the evaporation
of the black hole.  Using the formula of the Bekenstein-Hawking
entropy, the black hole mass is introduced to our model.  Then,
analysis was performed assuming that a single frequency mode passes
independently through the quantum circuit.  We revealed that the
negativity between \textsf{BH} and \textsf{JR} becomes zero after the
Page time for $M_4\omega\leq 0.041$ and the firewall-like structure
appears. This separable state has the small non-zero mutual
information, which becomes zero in $M_4\omega\rightarrow 0$ limit. On
the other hand, for $ M_4 \omega>0.041$, the entanglement does not
become zero and we do not have the firewall-like structure.  One
notice concerning our model is that it contains the structure of
horizon. In the burning paper model, particles' information is
released to the outside as the ash, hence it is a model without
horizon. In the model introduced in this paper, \textsf{BH} qubits are
discharged to the outside through CNOT-U gate, which squeezes inputted
qubits states and mimics the Hawking radiation. This means the effect
of the horizon is taken into our model.  Another notice is that we
assume $\ket{0}_\textsf{BH}$ as the initial state of \textsf{BH}. It
may be possible to adopt the initial state with some excitation of
particles determined by temperature of the initial mass of the black
hole. However, we expect that such a generalization does not alter the
main feature of entanglement obtained in this paper because the
non-vacuum weakly entangled initial state of \textsf{BH} will not
change the entanglement structure after the Page time so much.

Related to our investigation in this paper, the result of
S.~Luo \textit{et al.}  \cite{Albrecht_multi} may correspond to the
high frequency mode of our model and they found that the firewall is
not necessary. However, our conclusion is that by considering low
frequency modes, the black hole always shows the firewall behavior
after the Page time.  AMPS pointed out necessity of some mechanism
to affect the function of horizon to keep the theoretical consistency.
Our result can be interpreted that the long wavelength mode influences
the horizon so as to form the firewall.  In particular, for the
  low frequency mode with $ M_4 \omega\leq 0.041$, the Hawking
  particles emitted earlier than the Page time influences the function
  of the horizon after the Page time. They form the GHZ state with
  newly created Hawking particle pairs.  Therefore, \textsf{BH} and
  \textsf{JR} can only possess the classical correlation without
  entanglement.

  The impact of the low frequency mode (soft mode) of quantum
  fluctuations on the entanglement structure is also discussed in
  different contexts and different systems.  In the
  paper~\cite{Hotta2018}, if the zero-energy soft mode emission is
  involved with the evaporation process, the entanglement entropy
  between the black hole and the radiation becomes much larger than
  the black hole's thermal entropy as opposed to the Page curve
  prediction.  In cosmological situations, soft modes of which
  wavelength is larger than the Hubble horizon scale, becomes
  separable and shows the similar entanglement structure to the
  firewall~\cite{Matsumura2018}.  It may be interesting to obtain
  unified understanding of emergence of the separable state from the
  viewpoint of the horizon structure (geometry) and the monogamy
  property (entanglement).



\end{document}